\documentclass[conference]{IEEEtran}
\usepackage{graphicx} 

\title{Lock-free Asynchronously Distributed Linked Lists}

\author{\IEEEauthorblockN{Raaghav Ravishankar}
\IEEEauthorblockA{
\textit{Michigan State University}\\
East Lansing, MI, USA \\
ravisha7@msu.edu}
\and
\IEEEauthorblockN{Sandeep Kulkarni}
\IEEEauthorblockA{
\textit{Michigan State University}\\
East Lansing, MI, USA \\
sandeep@msu.edu}
\and
\IEEEauthorblockN{Sathya Peri}
\IEEEauthorblockA{
\textit{Indian Institute of Technology}\\
Hyderabad, Telangana, India \\
sathya\_p@cse.iith.ac.in}
\and
\IEEEauthorblockN{Gokarna Sharma}
\IEEEauthorblockA{
\textit{Kent State University}\\
Kent, OH, USA \\
gsharma2@kent.edu}
}

\usepackage{subfiles} 

\usepackage[inline]{enumitem}
\setlist[enumerate,1]{label={(\arabic*)}}

\usepackage{multicol}
\usepackage{xspace}
\usepackage{subcaption}
\usepackage{comment}
\usepackage{color}
\usepackage{setspace}
\usepackage[plain, linesnumbered, noend, noresetcount]{algorithm2e}
\makeatletter
\renewcommand{\@algocf@capt@plain}{above}
\renewcommand{\algocf@caption@plain}{\box\algocf@capbox\vskip\AlCapSkip}%
\makeatother
\usepackage{paralist}
\usepackage{amsmath,amsthm, nccmath}

\usepackage{mathtools}
\usepackage{cleveref}
\usepackage{url}

\newtheorem{definition}{Definition}[section]

\newtheorem{theorem}{Theorem}[section]
\newtheorem{lemma}[theorem]{Lemma}

\SetNlSty{bfseries}{\color{black}}{}

\crefname{algocf}{alg.}{algs.}
\Crefname{algocf}{Algorithm}{Algorithms}
\SetKwRepeat{Do}{do}{while}
\SetKwProg{Fn}{}{:}{end}
\SetKwProg{Struct}{struct}{:}{}
\SetKwComment{Comment}{\slash \slash }{}
\SetKwIF{If}{ElseIf}{Else}{if}{}{else if}{else}{end if}

\definecolor{asparagus}{rgb}{0.53, 0.66, 0.42}

\newcommand{\raaghavnote}[1]{{\color{magenta}#1}}
\definecolor{highlight}{rgb}{0.54, 0.17, 0.89}

\newcommand{\subtail}{{\tt SubTail\xspace}\xspace}
\newcommand{\sublist}{{sublist}\xspace}
\newcommand{\sublists}{{sublists}\xspace}
\newcommand{\Registry}{{\tt Registry\xspace}\xspace}

\newcommand{\subhead}{{\tt SubHead\xspace}\xspace}

\newcommand{\ItemRef}{{\tt Ref\xspace}\xspace}
\newcommand{\Item}{{\tt Item\xspace}\xspace}
\newcommand{\Entry}{{\tt Entry\xspace}\xspace}

\newcommand{\InsertAfter}{{\tt{InsertAfter}}\xspace}
\newcommand{\Insert}{{\tt Insert}\xspace}

\newcommand{\Key}{{\tt Key}\xspace}
\newcommand{\Delete}{{\tt{Delete}}\xspace}
\newcommand{\Split}{{\tt Split}\xspace}
\newcommand{\Move}{{\tt Move}\xspace}

\newcommand{\newLocation}{{newLoc}\xspace}

\newcommand{\DistributedLockFreedomTitle}{{Conditional Lock-Freedom}\xspace}
\newcommand{\DistributedLockFreedom}{{Conditional lock-freedom}\xspace}

\newcommand{\distributedLockFreedom}{{conditional lock-freedom}\xspace}
\newcommand{\distributedlockfree}{{conditional lock-free}\xspace}

\newcommand{\StartCount}{{stCt}\xspace}

\newcommand{\EndCount}{{endCt}\xspace}
\newcommand{\offset}{{offset}\xspace}

\begin{document}



\maketitle
\begin{abstract}
Modern databases use dynamic search structures that store an enormous amount of data, and often serve them using multi-threaded algorithms to support the ever-increasing throughput needs. When this throughput need exceeds the capacity of the machine hosting the structure, one either needs to replace the underlying hardware (an option that is typically not viable and introduces a long down time) or make the data structure distributed. Static partitioning of the data structure for distribution is not desirable, as it is prone to uneven load distribution over time, and having to change the partitioning scheme later will require downtime. 

The goal of this paper is to extend a concurrent data structure to distributed data structures that provide dynamic load balancing while preserving important properties such as lock freedom. With this intuition, first, we introduce the notion of \distributedLockFreedom which extends the notion of lock-free computation with reasonable assumptions about communication between processes. Then, we present  DiLi, a \distributedlockfree, linearizable, and distributable linked list that can be asynchronously and dynamically (1) partitioned into multiple sublists and (2) load balanced by distributing sublists across multiple machines. DiLi contains primitives for these that also maintain the lock-free property of the underlying search structure that supports find, remove, and insert of a key as the client operations.

We show that DiLi bridges the gap between concurrent data structures and distributed data structures. Specifically, DiLi provides comparable (and better in write-intensive workloads) performance to skip lists (which are typically the fastest data structures for search in a concurrent environment). In addition, it provides horizontal scaling with dynamic load balancing in a distributed environment.

\end{abstract}





\section{Introduction} \label{sec:introduction}
    Concurrent dynamic search structures play a crucial role in many modern databases \cite{hbase,rocksdb,leveldb,SingleStore}. With the rising number of concurrent operations to a database, memory contention should be managed with care to preserve the atomicity of its operations. For this, locks are often utilized, which can lead to pitfalls such as deadlock, priority inversion, and convoying \cite{herlihybook}.
    
Lock-free search structures avoid the drawbacks of lock-based methods but are limited to single-machine shared memory systems. Scaling further will require either a hardware upgrade to the single machine and perform a data migration, or a scheme to distribute load across multiple machines, with each machine handling only a `part' of the data structure. The former can be expensive if the machine is self-hosted (for privacy reasons), or can be limited to the highest available instance if cloud-hosted. Migrations here would still cause downtime to the database. Further, the need to distribute also arises from the observation that Moore's law now has diminishing returns \cite{PostMooresLaw}.  Distribution to multiple machines is also necessary when the data structure by its own becomes huge, such as in social network and hyperlink graphs.



Partitioning for load balancing can be static (fixed key distribution per machine) or dynamic (key distribution evolves over time). Static schemes cannot adapt to workload shifts or scale without downtime. Dynamic schemes offer flexibility, but require shared memory structures to be temporarily unavailable during redistribution. We consider the question: Is it possible to partition and dynamically load balance a data structure while preserving the property of lock freedom. As an instance, we focus on the list data structure. 
Specifically,
we focus on \distributedlockfree (that extends lock-freedom to distributed environments) linked lists. Linked lists are fundamental data structures that can in turn construct other data structures \cite{Valois}. Specifically, we distribute the well-known Harris Linked list \cite{harris2001pragmatic} and improve its performance even on a single-machine setup. We show that the resulting list, DiLi, can also replace skip lists in write-intensive workloads, thus offering both superior performance and  distributability.
Due to the distributed scalability and the general performance improvement, any multi-threaded lock-free data structure implemented with linked lists as its foundation (e.g., graphs built from adjacency lists\cite{GraphDBSurvey}, sharding in blockchains\cite{shardingInBlockChains,shardingInBlockchains2, ShardingInBlockchains3}), can automatically become faster and distributable by replacing the linked list with DiLi for its foundation.   

\noindent
\textbf{Contributions:} 
We introduce DiLi, a distributed linked list that performs \emph{find, insert, and remove} of a key both in a multi-threaded and a multi-machine setup while supporting asynchronous, dynamic re-partitioning. Any part of this data structure (henceforth called a sublist) can be \textit{split} into two smaller parts. 


By splitting the data structure into sublists, each sublist can be individually \textit{moved} to another machine, thus \textit{switching} the load of this sublist to the other machine. Such a dynamic load distribution is challenging to implement asynchronously. We provide an interface to implement any dynamic load balancing scheme using our ``Split'' and ``Move'' operations. We achieve this with only  single-word compare-and-swap (CAS) and fetch-and-add atomic instructions, which are now available in most commodity hardware. 


We design DiLi for write-heavy workloads and, as already investigated in synchrobench \cite{synchrobench}, lock-free skip lists outperform lock-free trees in write-heavy workloads due to the lack of rebalancing overhead. Hence, we compare DiLi with a popular lock-free skip list\cite{FraserDissertation} (all references to skip list in this paper refer to  \cite{FraserDissertation}). We show that
\begin{enumerate*}
    \item The performance of DiLi is not only comparable to lock-free skip lists, but even faster in write-intensive workloads. 
    \item DiLi is capable of automatic scaling up to several machines with a linear growth in throughput, and
    \item The performance of Split and Move operations are practical even under write-intensive workloads. 
\end{enumerate*}

\noindent
\textbf{Outline:} 
We introduce the notion of \distributedLockFreedom in \Cref{sec:DistributedLockFreedom},  our system model in \Cref{sec:systemModel}, and explain our design principles in \Cref{sec:designPrinciples}. DiLi is then described through the following three sections: Algorithm (\Cref{sec:Dili}), Correctness (\Cref{sec:Correctness}) and Empirical Evaluation (\Cref{sec:Empirical Evaluation}). Then we compare DiLi with related work in \Cref{sec:relatedWork}. Finally, the concluding remarks are in \Cref{sec:Conclusion}.

\section{Lock-Freedom for Distributed Systems}
\label{sec:DistributedLockFreedom}
    Even in simple client-server models such as a single server-client model, the \textit{traditional} lock-freedom definition: ``A method is lock-free if it guarantees that infinitely often some method call finishes in a finite number of steps'' \cite{herlihybook} is difficult to achieve, as any client request has to get past unbounded network delays to get the operation request on a data structure served. Also, the client thread that initiates this request typically sleeps until another thread responds back to it. 
    Due to the above constraints, we condition any communication between threads (such as messages) in the distributed system to only take a finite number of steps (e.g., with reliable channels). 

\begin{definition}[\DistributedLockFreedomTitle]
    We say that a method (or operation) implementation is \distributedlockfree if it is traditional lock-free when the execution satisfies the following condition: any communication between threads (such as messages) in the distributed system only takes a finite number of steps (e.g., with reliable channels).
    
\end{definition}

We have chosen to define \distributedLockFreedom in this fashion to distinguish between a simple scenario where a server in a client-server model (1) implements the Harris list, and (2) implements a traditional lock based list. The former will be \distributedlockfree while the latter would not.

\section{System Model}
\label{sec:systemModel}
\begin{figure}
    \centering
    \includegraphics[scale=0.28]{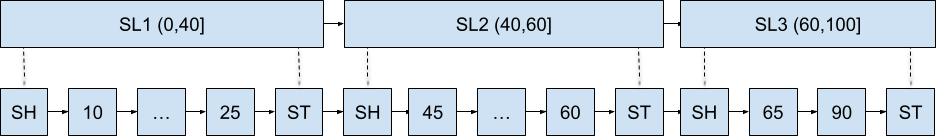}
    \caption{A sample list for keys 1--100 being composed of a set of 3 \sublists.}
    \label{fig:archSublists}
\end{figure}

    We study a list sorted by key and refer to each of its elements as \textit{items}. To construct our partitioned version of this list, we introduce the notion of a sublist that comprises a set of items contained by a \textit{subhead} (SH) and a \textit{subtail} (ST). 
    For example, in \Cref{fig:archSublists}, we show a list consisting of 3 sublists (SL1 to SL3).
    The list supports the following three operations:
    \begin{enumerate*}[itemsep=0mm, topsep=0pt]
        \item Find(key): Returns false if the key is not found in the list and otherwise returns true.
        \item Insert(key): Returns false if the key is already in the list and otherwise inserts the key in its sorted position, returning true.
        \item Remove(key): Returns false if the key is not found in the list, and otherwise deletes the key, de-links the item off the list, and returns true.  
    \end{enumerate*}
    Additionally, a server may invoke operations Split (to partition the sublist into two sublists), Move (to create a live clone of the sublist in another server), and Switch (to transfer ownership of a sublist to a new server). We also support a Merge operation (delegated to Appendix \ref{sec:appendixMerge} for space) for two adjacent sublists. 

    We consider a client-server model of a distributed system with any number of multi-core machines acting as servers. The server and client machines are fully connected by reliable channels that guarantee eventual delivery. All messages exchanged in the channels are via remote procedure calls (RPCs).
    A server may not invoke any RPC on the clients.
    However, servers may invoke RPCs on each other and also send messages to each other asynchronously.

\begin{figure}
    \centering
    \includegraphics[scale=0.35]{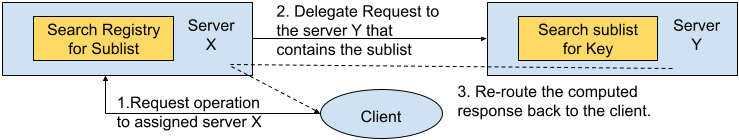}
    \caption{Client invokes RPC on a key to its assigned Server X and X delegates the request to Server Y that owns the key.}
    \label{fig:clientServerModel}
\vspace*{-3mm}
\end{figure}

    As in \Cref{fig:clientServerModel}, we follow a decentralized architecture. Each client will contact a pre-assigned server X who will, if necessary, \textit{act as a proxy} and \textit{delegate} the request to server Y if Y holds the key $k$ used by the client operation. 
    Note that X could also receive delegated requests from other servers as it also owns some sublists. 
    We assume that a machine does not crash entirely but has at least one core. This assumption is essential, as each item is stored in only one server.

    
    A concern when aiming for linearizability\cite{linearizability} in a distributed system is to operate under the constraints of the CAP theorem \cite{CAP}.
    By ensuring that any operation takes place on only one sublist, and that a sublist entirely resides in one single machine at any given time, we are able to overcome the effects of network partitioning that impedes linearizability.
    We 
    allow a sublist to undergo a Split, Move, or Switch by only one thread at a time. This can be easily achieved by assigning only one background thread per machine to perform such operations. 
    Note that even if we have at most one background thread operating on a sublist at any time, we are required to ensure the lock-free nature of the client operations on the list. We refer to this guarantee as \textit{the asynchronous nature} of the background operations.    

\section{Design Principles}
\label{sec:designPrinciples}
    The design of the distributed list involves several techniques in order to provide a linearizable lock-free implementation. The first is a \textbf{smart pointer}, where the least significant bit (LSB), which is normally 0 for address alignment, is used to denote if an item has been soft deleted.
In addition, there is a common
    limitation on the
    x86-64 MMUs of commodity hardware (such as without 5-level paging extensions \cite{intel5LevelPaging}) only to have a 48-bit virtual address space, leaving 16 most significant bits unused. 
    DiLi uses these 16 bits to globally indicate the server ID that the distributed list item belongs to. For futuristic hardware that supports larger pointer values, this allocation can be adjusted accordingly.


    Secondly, we maintain all sublists as \textit{entries} in a lazily replicated \textit{registry}—an index where each entry has a smart subhead pointer that encodes the server ID. This ID identifies the server that hosts the sublist and guides request routing. Although lazy replication may cause temporary staleness and latency, our safeguards (see \Cref{sec:Dili}) ensure that lock-freedom and linearizability remain intact.
    Lazy replication allows the registry to remain sorted via copy-on-write inserts. This enables logarithmic-time key lookup. 
    This leads to our next design principle, \textbf{the hybrid search}, where we use a logarithmic search in the registry followed by a linear search in the sublist.
    By keeping a check on the number of items present in a sublist, we achieve quick traversals in the distributed list. 


    The Move operation handles the load balancing by creating
    a momentary live clone of the sublist in the new machine. To achieve this in a non-blocking manner\footnote{Note that it is insufficient to use snapshotting algorithms for this purpose, because we need a snapshot that always matches the live state of the sublist, instead of a state at some point of time in the past.}, we perform \textbf{temporary replication} of updates that happen during the move. 
    For instance, if insert($k$), remove($k$), and insert($k$) occur in order, both replicas must retain $k$ - avoiding out-of-order replay like insert($k$), insert($k$) and remove($k$). This challenge stems from mismatches between the linearized update order and the delivery order. We achieve this with our  \textbf{novel replay algorithm} using logical timestamps.

    When Move  terminates, the cloned sublist is activated, and the ownership of the sublist is transferred to the new server by Switch operation. The Switch operation lazily updates the registry on all servers. Careful delegations are used to handle client operations when the registry information is temporarily outdated. 
    
    Although blocking client operations can simplify this transfer, we achieve it asynchronously using integer counters around sublist updates.

\section{D\lowercase{i}L\lowercase{i}: A \underline{D\lowercase{i}}\lowercase{stributable} \underline{L\lowercase{i}}\lowercase{{st}}}
\label{sec:Dili}
DiLi is a linearizable sorted linked list with a two-layered data structure (described in \Cref{subsec:dataStructures})  -- (1) an index called the Registry, that keeps track of the subhead and subtail pointers (to shortcut into a sublist), and (2) sublists that consist of the underlying list items. Upon reaching a subhead, a linear traversal is employed to find the search node. The find, insert, and remove operations (as described in \Cref{subsec:clientOps}) that utilize this traversal are lock-free,
but only operate on the sublist space. For ease of explanation, we use colors in the algorithms. Black lines are executed the same way, regardless of whether there is an ongoing background operation, the red lines adapt to a concurrent Split and hence will be explained as part of \Cref{subsec:split}, and the blue lines adapt to a concurrent Move/Switch and are explained in \Cref{subsec:MoveSwitch}.
The novel contribution in our approach is an asynchronous yet versatile Split, Move, and Switch algorithms that can be used to build any partitioning and load balancing scheme of choice to suit the workload of interest. 

\subsection{Data Structures: List Item and Registry}
\label{subsec:dataStructures}
\Cref{alg:Main} depicts the two building blocks of DiLi. Like any list item, DiLi items contain a key to search for the item and a next pointer to lead to the next item. Note that this is a distributed list, with a traversal guided by the next pointer. We denote the type of this pointer as $\ItemRef$. Any $\ItemRef$ pointer, say X, needs to be operated on by a bit mask to force the first 16 bits and the last bit to 0 before accessing the address that X points to. We overload this functionality into the C++ pointer access notation ($\rightarrow$). 
Fetching serverID bits, and if the item has been marked, are also implemented through bitmasks.
We denote them as X.id and X.mark in the algorithms.


The registry consists of sublist entries and supports a getByKey(key) operation to find the sublist entry to which a searched key belongs. Operation addEntry(entry) is used to insert an entry into an already sorted registry. We utilize a copy-on-write (COW) technique to update the registry pointer, to atomically update the registry with a new sorted registry with an added entry. Because registry replication is done lazily, it is sufficient to have only one update happening at a time on the registry. Hence any balanced single-writer tree data structure (binary tree, B-tree, or even a skip list) could also be used to build the registry, thereby offering a logarithmic search for the getByKey(key) operation. For the sake of simplicity, we maintain the registry as a binary search tree in a large array and keep track of its size.
We also utilize a variant of hazard pointers \cite{hazardPointers} to reclaim memory when an old pointer of the registry (after being overwritten by COW) is no longer referenced by any thread. To save space, the functioning of the registry is explained in Appendix \ref{sec:appendixRegistry}. We also reserve two keys out of the available key range for the list to indicate the subhead (SH\_KEY) and subtail (ST\_KEY) items.

\subsection{Client Operations}
\label{subsec:clientOps}
    Here, we present the functioning of Find, Remove, and Insert operations assuming that the partitioning index (registry) and the ownership of various sublists by different machines stay static. When that assumption breaks, the corresponding changes to their execution are explained in Sections \ref{subsec:split} (red lines) and \ref{subsec:MoveSwitch} (blue lines), respectively.

    All client operations perform the following in order: \begin{enumerate*}
        \item A binary search on the replicated registry to retrieve the sublist entry for a key.
        \item Routing the request to the appropriate server based on the server ID of the subhead present in the sublist entry.
        \item A linear traversal on the sublist to perform operations in a Harris-List fashion with retries beginning from the subhead of the sublist.
    \end{enumerate*}

    A registry look-up is first performed for any operation requested by the client. This returns the subhead entry for the key range, that in-turn contains the $\ItemRef$ for its subhead. By checking the id bits of $\ItemRef$, the machine serving the sublist is determined to reroute as needed. The operations take only the key as a parameter if the request comes directly from a client, and if it is instead a delegation, the subhead is also a parameter. A re-routed request already has the subhead required, and hence skips the registry lookup step on its machine. 

    The end machine that serves the sublist traverses using the Search algorithm in \Cref{alg:Main}. It resembles Harris List search, where the left and right nodes to the searched key are returned, while delinking any marked items encountered along the traversal. However, we optimize this search by limiting the search to start from the pre-determined subhead, instead of the list head. In our case, the rightNode is the item of the searched key or the one immediately greater present in the list. leftNode is the corresponding previous item. 
    
    Find and Remove utilize the returned right node of a search to determine if a key is present. Find returns its response based on this check (after Line \ref{Line:FindSearchResponse}), while Remove continues with a Delete on the found item using the Delete function. The Delete attempts to \textit{soft} delete the node by marking its next pointer's mark bit using CAS. If the CAS succeeds (Line \ref{Line:DeleteCAS}), then the Remove returns true, and if it is already marked, then it returns false.

    The Insert operation as described in \Cref{alg:Main} utilizes the InsertInSublist routine after determining the sublist to insert the key. This routine begins with a search operation to determine the left and right node window. If the right node matches the key to be inserted, then the Insert returns false. If the next pointer of left node is altered by getting marked or through a competing insert, then the CAS fails and the search is restarted on the same sublist to retry the insert. The logic behind the \StartCount, \EndCount, \newLocation and ts assignments becomes apparent only during the background operations, as covered in the next sections.
    

\begin{algorithm*}
\caption{Algorithm for $\tt{Find}$, $\Insert$, $\tt{Remove}$ and $\Split$ Operations}
\label{alg:Main}
\SetInd{0.25em}{0.5em}
\vspace*{-7mm}
\scriptsize
\begin{multicols}{3}
\textbf{Data structures used:
}
\vspace{3mm}
\begin{multicols}{2}
\Struct{$\Item$} {
    $\Key$ key\;
    \textcolor{red}{$\Key$ keyMax\;}
    \textcolor{blue}{int ts\;
    int sId\;}
    $\ItemRef$ next\;
    \textcolor{blue}{int \StartCount\;
    int \EndCount\;
    $\ItemRef$\ \newLocation\;}
}
\Struct{$\Entry$} {
    $\ItemRef$ subhead\;
    $\ItemRef$ subtail\;
    $\Key$ keyMin\;
    $\Key$ keyMax\;
    int \StartCount\;
    int \EndCount\;
    int offset\;
}
\end{multicols}
\Struct{$\Registry$} {
    $\Entry$\ignorespacesafterend* entries[MAX\_SUBLISTS] = $\{$null, $...$ , null$\}$\;
    int size = 0\;
}
\BlankLine
\vspace{10mm}
\vspace{-5mm}
\textbf{Algorithms:}
\vspace{5mm}

\Fn{\normalfont$\ItemRef$ Search($\Key$ key, $\ItemRef$ head, $\ItemRef$ \normalfont{\&}leftNode)}{
    curr = head\;

    \textcolor{blue}{
        \If{curr$\rightarrow$\StartCount $<$ 0}{
            return head$\rightarrow$\newLocation\;
        }
    }
    \While{true}{
        leftNode = prev = curr\;
        curr = curr$\rightarrow$next\;
        \textcolor{blue}{
        \If{curr$\rightarrow$\StartCount $<$ 0\label{line:searchStartCountNegativeCHeck}}{
            head = (registry.getByKey(key)) $\rightarrow$subhead\;
            \If{head.id == me}{
                leftNode = prev = head\;
                curr = prev$\rightarrow$next\;
                \If{curr$\rightarrow$\StartCount $<$ 0}{
                    return head$\rightarrow$\newLocation\;
                }
                continue\;
            }
            return head\;
        }}
        \If{curr$\rightarrow$key == ST\_KEY \label{Line:subtailCheckSt}}{
            \label{lines:searchSplitInteractionStart}
            \If{key $\leq$ curr$\rightarrow$keyMax \label{Line:searchLastSubtailCheck}}{
                return null\;
            }
            \textcolor{red}{
            curr = curr$\rightarrow$next\;
            \If{curr.id $\neq$ me}{
                return curr\;
            }
            leftNode = prev = curr\;
            head = curr\;
            curr = curr$\rightarrow$next\;
            }\label{Line:subtailCheckEnd}   
        }
        \If{curr$\rightarrow$next.mark \label{line:searchMarkCheck}}{
            \eIf{delinkNode(prev, curr)}{
                \textcolor{red}{Lines \ref{Line:subtailCheckSt} to \ref{Line:subtailCheckEnd}}\;
                continue\;
            } {
            leftNode $=$ prev $=$ head\;
            curr $=$ prev$\rightarrow$next\;
            \textcolor{blue}{
            \If{curr$\rightarrow$\StartCount $<$ 0}{
                return head$\rightarrow$\newLocation\; }
            }  
            continue\;
            }
        }
        \If{\normalfont curr$\rightarrow$key $==$ key $\wedge$ curr$\rightarrow$\StartCount $\geq$0 \label{line:SearchFoundReturn}}{
                return curr\;
        }
        \If{\normalfont curr$\rightarrow$key $>$ key $\wedge$ curr$\rightarrow$\StartCount $\geq$ 0 \label{line:Search404Return}}{
            return null\;
        }
    }
    \textcolor{blue}{
    \If{curr$\rightarrow$\StartCount $<$ 0}{
        return head$\rightarrow$\newLocation\;
    }
    }
    return null\;
}
\BlankLine
\Fn{\normalfont bool delinkNode($\ItemRef$ prev, $\ItemRef$ \normalfont{\&}curr)} {
    currNext = curr$\rightarrow$next\;
    \While{currNext.mark == 1}{
    currNext = currNext$\rightarrow$next\;
    }
    temp = curr\;
    \If{CAS(prev$\rightarrow$next,temp,currNext)\label{line:delinkCAS}} {
        curr = currNext\;
        return true\;
    } 
    return false\;
    
}
\BlankLine

\Fn{\normalfont bool $\tt{Find}$($\Key$ key, $\ItemRef$ SH = null)} {
\If{SH == null $\lor$ SH$\rightarrow$\StartCount $<$ 0}{
    entry = registry.getByKey(key)\;
    entrySH = entry$\rightarrow$subhead\;
}

sId = SH.id\;
\If{sId == me}{
    leftNode = null\;
    node = Search(key, SH, leftNode)\;\label{Line:FindSearchResponse}
    \If{node == null}{
        return false\;
    }
    sId = node.id\;
    \If{sId == me}{
        return true\;\label{line:findsuccess}
    }
    \textcolor{blue}{
        response = Send Find(key,node) to sId\;
        return response\;
    }
}
response = Send Find(key,SH) to sId\;
return response\;
}
\BlankLine
\Fn{\normalfont bool $\Insert$($\Key$ key, $\ItemRef$ SH = null)}{
\If{SH == null $\lor$ SH$\rightarrow$\StartCount $<$ 0}{
    entry = registry.getByKey(key)\;
    entrySH = entry$\rightarrow$subhead\;
}

sId = SH.id\;
\If{sId == me}{
    return InsertInSublist(key, SH)\;
}
response = Send Insert(key,SH) to sId\;
return response\;
}
\BlankLine
\Fn{\normalfont bool InsertInSublist($\Key$ key, $\ItemRef$ SH)}{
\While{true}{
    leftNode = null\;
    rightNode = Search(key, SH, leftNode)\;
    \If{rightNode $\neq$ null}{
        \If{rightNode.id == me}{
            return false\;
        }
        response = Send Insert(key,rightNode) to rightNode.id\;
        return response\;
    }

    temp = leftNode$\rightarrow$next\;
    \If{temp$\rightarrow$key == key}{
        \If{temp$\rightarrow$\StartCount $\geq$ 0}{
            return false\;
        }
        \textcolor{blue}{
        \label{line:insertDelegateSt}
        SH = (registry.getByKey(key)) $\rightarrow$subhead\;
        \If{SH.id == me}{
            continue\;
        }
        response = Send Insert(key, SH$\rightarrow$\newLocation)\;
        return response\;
        \label{line:insertDelegateEnd}
        }
    }
    \textcolor{red}{
    \If{temp$\rightarrow$key == ST\_KEY}{
        \If{temp$\rightarrow$keyMax $<$ key}{
            SH = temp$\rightarrow$next\;
            continue\;
        }
    }}
    leftNode$\rightarrow$\StartCount$\rightarrow$fetch\_add(1)\;
    \textcolor{blue}{
    \If{leftNode$\rightarrow$\StartCount $<$ 0}{
        Lines \ref{line:insertDelegateSt} to \ref{line:insertDelegateEnd}.
    }
    }
    newItem = new Item(key, $\phi$, ts.fetch\_add(1), me, temp, leftNode$\rightarrow$\StartCount, leftNode$\rightarrow$\EndCount,
    leftNode$\rightarrow$\newLocation)\;
    \If{CAS(leftNode$\rightarrow$next, temp, $\ItemRef$(newItem)\label{line:InsertCAS}}{
    \eIf{newItem$\rightarrow$\newLocation $\neq$ null}{
        \textcolor{blue}{
        Send RepInsertAfter(leftNode$\rightarrow$ts, leftNode$\rightarrow$sId, leftNode$\rightarrow$\newLocation, newItem$\rightarrow$ts, newItem$\rightarrow$sId, newItem, key) to newItem$\rightarrow$\newLocation.id\;\label{line:repInsertCall}
        }
    } {
        leftNode$\rightarrow$\EndCount$\rightarrow$fetch\_add(1)\;
    }
    return true\;
    }
 leftNode$\rightarrow$\EndCount$\rightarrow$fetch\_add(1)\;
}
}
\BlankLine
\Fn{\normalfont bool $\tt{Remove}$(Key key, $\ItemRef$ SH = null)}{
    Same as Find but line \ref{line:findsuccess} is changed to
    
    return Delete(node, key)\; and the corresponding delegates are that of $\tt{Remove}$ instead of $\tt{Find}$.  
}
\BlankLine
\Fn{\normalfont bool Delete($\ItemRef$ node, Key key)} {
    result = false\;
    \If{node$\rightarrow$next.mark}{
            return false\;
    }
    node$\rightarrow\StartCount\rightarrow$fetch\_add(1)\;

    \textcolor{blue}{
    \If{node$\rightarrow$\StartCount $<$ 0}{
        response = Send Delete(node$\rightarrow$\newLocation,key)\; 
        return response\;
    }
    }
    \While{true}{
        \If{node$\rightarrow$next.mark\label{line:itemAlreadydeleted}} {
            node$\rightarrow$\EndCount$\rightarrow$fetch\_add(1)\;
            break\;
        }
        temp = node$\rightarrow$next\;
        newVal = temp\;
        newVal.mark = true\;
        \If{CAS(node$\rightarrow$next, temp, newVal)\label{Line:DeleteCAS}}{
            result = true\;
            \eIf{node$\rightarrow$\newLocation $\neq$ null}{
                \textcolor{blue}{
                Send RepDelete (node$\rightarrow$\newLocation, node, node$\rightarrow$ts, \\ node$\rightarrow$sId) to node$\rightarrow$\newLocation.id\;
                }
            } {
                node$\rightarrow$\EndCount$\rightarrow$fetch\_add(1)\;
            }
            break\;
        }
    }
    leftNode = null\;
    SH = registry.getByKey(key)$\rightarrow$subhead\;
    Search(key, SH, leftNode)\;
    return result\;
}
\BlankLine
\Fn{\normalfont $\Entry$* $\Split$($\Entry$* entry, $\ItemRef$ sItem)} {
    *new\StartCount = new int[1]\;\label{line:startcount}
    *new\EndCount = new int[1]\;\label{line:endcount}
    SH = new Item(SH\_KEY, $\phi$, ts.fetch\_add(1), me, null, \\ new\StartCount, new\EndCount, null)\;\label{line:splitSTCreate}
    ST = new Item(ST\_KEY, sItem$\rightarrow$key, ts.fetch\_add(1), me, SH, sItem$\rightarrow$\StartCount, sItem$\rightarrow$\EndCount, null)\;\label{line:splitSHCreate}

    \Do{
        $\neg$CAS(sItem$\rightarrow$next, temp, ST)  \label{line:splitInsertion}  
    }{
    temp = sItem$\rightarrow$next\;
        \If{temp.mark}{
            return null\;\label{line:splitFails}
        }
        SH$\rightarrow$next = temp\;
        SH$\rightarrow$ts = ts.fetch\_add(1)\;
    }

    curr = SH$\rightarrow$next\;
    \Do{
        prev$\rightarrow$key $\neq$ ST\_KEY \label{line:splitCountUpdateEnd}
    }{
        \label{line:splitCountUpdateStart}
        prev = curr\;
        curr$\rightarrow$\StartCount = new\StartCount\;
        curr$\rightarrow$\EndCount = new\EndCount\;
        curr = curr$\rightarrow$next\;
    }
    \Do{$a_1 + a_2 \neq$ entry$\rightarrow$ offset \label{line:splitOffsetEnd}}{
            \label{line:splitOffsetStart}

        $a_1$ = new\StartCount - new\EndCount\;
        $a_2$ = sItem$\rightarrow$\StartCount - sItem$\rightarrow$\EndCount\;
    } 

    newEntry = new $\Entry$(SH, prev, sItem$\rightarrow$key, entry$\rightarrow$keyMax, new\StartCount, new\EndCount, $a_1$)\;\label{line:splitRegistryStart}

    registry.addEntry(newEntry)\;
    entry$\rightarrow$\offset = $a_2$\;
    entry$\rightarrow$keyMax = sItem$\rightarrow$key\;
    entry$\rightarrow$subtail = ST\;

    \For{$i \in$ (serverList $-$ \{me\})}{
        response = Send RegisterSublist(sItem$\rightarrow$key, SH)\;\label{line:splitRegistryEnd}
    }
    
    return newEntry\;
}
\BlankLine
\Fn{\normalfont bool RegisterSublistRecv(Key keyMin, $\ItemRef$ SH)} {
    $\Entry$ leftEntry = registry.getByKey(keyMin)\;
    $\Entry$ rightEntry = new $\Entry$(SH, null, keyMin,\\ leftEntry$\rightarrow$keyMax, null, null, 0)\;
    leftEntry$\rightarrow$keyMax = keyMin\;
    return true\;
}
\BlankLine
\end{multicols}
\vspace*{-5mm}
\end{algorithm*}
\subsection{The Split Algorithm}
\label{subsec:split}
    The Split operation (\Cref{alg:Main}) takes two inputs -- a sublist entry and an item (referred to as sItem) at which the sublist is to be split. 
    This truncates the original sublist to a smaller key range and creates a new sublist for the remaining key range. 
    The Split is performed through 4 major steps, while always guaranteeing that the client operations stay lock-free: 
    \begin{enumerate*}
        \item Allocate new counters \StartCount and \EndCount (lines \ref{line:startcount} and \ref{line:endcount}) for the new sublist. The purpose of these counts is to ensure that if a client operation overlaps with Split/Move, it is detected so that Split/Move can take corrective action.
        \item Insert a connected block of a subtail followed by a subhead (as in Lines \ref{line:splitSTCreate} and \ref{line:splitSHCreate}) at the next pointer of sItem (insertion is in Line \ref{line:splitInsertion}). If sItem gets deleted before this, split operation fails (Line \ref{line:splitFails}) and can be retried with another call at a different sItem. 
        \item Update the \StartCount and \EndCount pointers of the second half of the sublist (Lines \ref{line:splitCountUpdateStart} to \ref{line:splitCountUpdateEnd}). Note that if an item gets inserted and gets missed in the traversal, it must have copied the counters from the item it was previous to, whose counters (or their ancestors') were updated during this traversal. Thus, they must also contain the new counter pointers.
        \item Update the registry by adding an entry for the second half of the sublist, and then truncating the first half to its reduced keyMax (Lines \ref{line:splitRegistryStart} to \ref{line:splitRegistryEnd}). 
    \end{enumerate*} 

    The above approach has two corner cases to handle due to the asynchronous insertion. First, for client operations that concurrently search for a key higher than sItem, the search can get truncated at the inserted subtail in case the client operation traversal encounters it. For this purpose, the subtails contain a keyMax element stored. Specifically, the red lines from \ref{lines:searchSplitInteractionStart} to \ref{Line:subtailCheckEnd} check whether the key belongs to the first half of the sublist that is undergoing a concurrent split or to the second half. If it is in the second half, then its subhead pointer is updated, speeding up potential retries of the lock-free Harris List traversal.

    Second, we can no longer expect the difference between \StartCount and \EndCount of the individual sublists to be 0 when there are no ongoing updates (insert/remove). This is because, during the counter updates to the second half of the sublist, some updates could have incremented the \StartCount of the left half and updated the \EndCount of the right half. This accumulated change to the counter difference when the sublist is static is denoted as the \textit{offset} of the sublist. As would be explained later, the new offset is important for Move and Switch operations. To compute this new offset, we employ a spin (Lines \ref{line:splitOffsetStart} to \ref{line:splitOffsetEnd}) that breaks only when the computed values $a_1$ and $a_2$ reflect the new right half offset and left half offsets respectively. The condition works because the difference between the \StartCount and \EndCount of any sublist stabilizes to a minimum only when there are no ongoing updates in the sublist. This is true for both the left and right halves of the sublist, and since the increments went to exactly one of the two \StartCount counters and one of the two \EndCount counters for each update, the sum of the new offsets must be equal to the old offset when the two new sublists together have no ongoing updates.  We note that this spin does not delay the client operations, and only delays background operations on the split sublists.

\subsection{Move, Replay and Switch Algorithms}
    A Move routine invoked on a server machine (henceforth $S_1$) takes the registry entry of a sublist and the machine to move it to (henceforth $S_2$), as parameters. The goal of Move is to create a clone of a sublist and then make it the live copy. To achieve this, the Move routine as described in \Cref{alg:MoveAndReplay} traverses each item (by recursively calling the MoveItem routine) in the sublist in $S_1$ and creates a copy of them in $S_2$. The reference in $S_2$ of last moved item is sent back each time so that subsequent items can be linked to the previously moved item.
    
    Move compensates for concurrent updates that the Move routine did not witness during its traversal, by \begin{enumerate*}    
     \item detecting that such concurrent updates occurred, and 
     \item \textit{replicating} those updates \textit{correctly} on $S_2$.
     \end{enumerate*}
     The former is achieved by using \StartCount (incremented before the start of update operation)  and \EndCount (incremented after the update operation); and the latter is achieved by sending replicates (Line \ref{line:repDeleteCall} in \Cref{alg:MoveAndReplay} and Line \ref{line:repInsertCall} in \Cref{alg:Main}) of these operations to $S_2$. As discussed in \Cref{sec:designPrinciples}, the inserts and removes will need further specification for the correct replay of the updates in $S_2$, as they may appear out of order when received. To account for this, the replay models the changes to the sublist as a result of the following update operations: InsertAfter($\ItemRef$, Item) and Delete($\ItemRef$), that distinctly convey where the insert CAS (Line \ref{line:InsertCAS}) and remove CAS (Line \ref{Line:DeleteCAS}) that had succeeded. If multiple items got inserted at a particular $\ItemRef$ $A$, that had been moved by the Move routine, then the item with the lower timestamp got inserted first. As for the Delete, the same item that was moved to $S_2$ is the one that gets deleted, and a subsequent insertion for the same key is treated as a different item. Since $\ItemRef$ present in $S_1$ for an item is not known by $\ItemRef$ copy of it in $S_2$, we denote the uniqueness of an item by the $<$sId, timestamp$>$ tuple.   

    Using the above observations, we developed our Replay algorithm presented in Lines \ref{Line:ReplayStart} to \ref{Line:replayEnd}. The received move of an item and a received insert replicate utilize this routine. Once the prev item is identified using their $<$sId, timestamp$>$ tuple, the Replay algorithm takes this prev item and traverses from prev to find the right position to insert the item. To not miss any item, the Move routine even moves marked items as part of the algorithm, which are then delinked once the cloned sublist becomes active. Because replay is concurrently invoked by Move and multiple Replicate threads, insertions at replay are retried. An extensive proof of correctness for the replay is provided in Appendix \ref{sec:AppendixReplay}. 

    Finally, to detect that a sublist is live in $S_1$, we utilize the \StartCount and \EndCount values of the sublist. All updates increment the \StartCount before performing the update, and increment the \EndCount after performing the update. However, if a replicate message is sent, the \EndCount is incremented only after the completion of the replay (Line \ref{line:replayInsertEndCt} and \ref{line:replayRemoveEndCt}). This way, when the \EndCount measured matches the \StartCount measured immediately after, we can be sure that both copies of the sublist are at the same state. At this point we use a CAS to set the \StartCount to $-\infty$ (Line \ref{line:minusiftyCAS}). All updates check if the \StartCount is negative after they increment it. If it is negative, they delegate the request to $S_2$. Otherwise, they continue with the operation, delaying the Move until the update increments \EndCount and so never block client operations.
    
    After \StartCount is permanently negative (the blue lines other than send replicates reflect this scenario), all updates are redirected towards $S_2$, thereby making its sublist version live and leaving the sublist at $S_1$ stale. This point begins the Switch operation (or phase). The Switch routine in \Cref{alg:MoveAndReplay} is invoked to first complete the re-distribution by updating the next pointer of the previous subtail. Since the current Switch has no control over the background operations on the sublist previous to it, this attempt is done in a loop to account for redirections (expected to not happen more than once, as it is highly undesirable to move the same sublist too frequently). After this step, the Switch, and hence the re-distribution is completed by updating the registry entry for the sublist in all servers.


\label{subsec:MoveSwitch}
\begin{algorithm*}
\caption{The Move and Replay Algorithm}  
\label{alg:MoveAndReplay}
\vspace{-7mm}
\SetInd{0.25em}{0.5em}
\scriptsize
\begin{multicols}{3}
\Fn{$\Move$($\Entry$ entry, int newSId)}{
head = entry$\rightarrow$subhead\;
remoteSH = Send MoveSH(head$\rightarrow$sId, head$\rightarrow$ts, entry$\rightarrow$keyMax)\;
head$\rightarrow$\newLocation = remoteSH\;
MoveNext(remoteSH, head$\rightarrow$next, newSId)\;
\Do{$\neg$CAS(entry$\rightarrow$\StartCount, temp, $-\infty$)\label{line:minusiftyCAS}}{
temp = entry$\rightarrow$\EndCount + entry$\rightarrow$offset\;
}
Switch(entry, newSId)\;
}
\Fn{\normalfont MoveNext($\ItemRef$ prev, $\ItemRef$ curr, int newSId)}{
\If{curr$\rightarrow$\newLocation == null}{
    temp = curr$\rightarrow$.mark\;
    curr$\rightarrow$\newLocation = Send MoveItem(curr$\rightarrow$key, prev, curr$\rightarrow$.mark, curr$\rightarrow$next,curr$\rightarrow$sId, curr$\rightarrow$ts) to newSId\;
}
\If{curr$\rightarrow$next.mark $\neq$ temp}{
    Send RepDelete(curr$\rightarrow$\newLocation, curr, curr$\rightarrow$ts, curr$\rightarrow$sId) to newSId\;\label{line:repDeleteCall}
}
\If{curr$\rightarrow$key == ST\_KEY}{
    return\;
}
MoveNext(curr$\rightarrow$\newLocation, curr$\rightarrow$next, newSId)\;
}

\Fn{\normalfont $\ItemRef$ MoveSHRecv(int itemSId, int itemTs, $\Key$ keyMax)}{
    *new\StartCount = new int[1]\;
    *new\EndCount = new int[1]\;
    ST = new Item(ST\_KEY, keyMax, , , null, new\StartCount, new\EndCount, null)\;
    SH = new Item(SH\_KEY, $\phi$, itemTs, itemSId, ST, new\StartCount, new\EndCount, null)\;
    entry = registry.getByKey(keyMax)\;
    entry$\rightarrow$subtail = ST\;
    entry$\rightarrow$offset = 0\;
    entry$\rightarrow$\StartCount = new\StartCount\;
    entry$\rightarrow$\EndCount = new\EndCount\;
    return SH\;
}

\Fn{\normalfont $\langle$$\ItemRef$, $\ItemRef$$\rangle$ RepInsertRecv($\ItemRef$ prevNewLoc,$\Key$ key, \ \ \  int prevSid, int prevTs, int itemSId, int itemTs)}{
    \While{prev$\rightarrow$ts $\neq$ prevTs $\lor$ prev$\rightarrow$sId $\neq$ prevSId     \label{Line:replaypositionstart}
}{
        prev = prev$\rightarrow$next\;
        \If{prev$\rightarrow$key == ST\_KEY}{
            prev = prevNewLoc\;
            
        }
        \label{line:replaypositionend}
    }
    return $<$prevNewLoc, Replay(prev, itemTs, key, itemSId, itemTs, false)$>$\;   
}
\Fn{\normalfont $\ItemRef$ RepDeleteRecv($\ItemRef$ prevNewLoc, int prevSId, int prevTs)}{
    Lines \ref{Line:replaypositionstart} to \ref{line:replaypositionend}\;
    \Do{$\neg$ CAS(prev$\rightarrow$next, temp, ref)}{
        temp = prev$\rightarrow$next\;
        ref = temp\;
        ref.mark = true\;
    }
    return prevNewLoc\;
}
\Fn{\normalfont $\ItemRef$ MoveItemRecv($\ItemRef$ prev, $\Key$ key, bool isMarked, $\ItemRef$ STNext, int itemSid, int itemTs)}{
\If{key == ST\_KEY}{
    curr = prev\;
    \Do{
    curr$\rightarrow$key $\neq$ ST\_KEY
    }{
    curr == curr$\rightarrow$next\;
    }
    curr$\rightarrow$next = STNext\;
    return curr\;
}
return Replay(prev, prev$\rightarrow$ts, key, itemSId, itemTs, isMarked)\;
}

\Fn{\normalfont $\ItemRef$ Replay($\ItemRef$ prev, $\ItemRef$ compTs, $\Key$ key, int itemSId, int itemTs, bool isMarked\label{Line:ReplayStart})}{
\Do{\normalfont $\neg$CAS(currPrev$\rightarrow$next, temp, ref)}{
    curr = prev\;
    \Do{
    curr$\rightarrow$ts $\geq$ compTs $\wedge$ curr$\rightarrow$key $\neq$ ST\_KEY
    \label{line:replayTsCompare}}{
        currPrev = curr\;
        curr = curr$\rightarrow$next\;\label{line:replayFindCurr}
    }

    temp = currPrev$\rightarrow$next\;
    newItem = new Item(key, $\phi$, itemTs, itemSId, temp, \\ \ \  currPrev$\rightarrow$\StartCount, currPrev$\rightarrow$\EndCount,
    currPrev$\rightarrow$\newLocation)\;
    newItem$\rightarrow$next.mark = isMarked\;
    ref = newItem\;
    ref.mark = temp.mark\;
}
return newItem\;    \label{Line:replayEnd}
}
\Fn{\normalfont InsertReplayResponseRecv($\ItemRef$ oldLoc, $\ItemRef$ newLoc)}{
    oldLoc$\rightarrow$\newLocation = newLoc\;
    oldLoc$\rightarrow$\EndCount$\rightarrow$fetch\_add(1)\;\label{line:replayInsertEndCt}
}
\Fn{\normalfont RemoveReplayResponseRecv($\ItemRef$ oldLoc)}{
    oldLoc$\rightarrow$\EndCount$\rightarrow$fetch\_add(1)\;\label{line:replayRemoveEndCt}
}

    \Fn{\normalfont $\tt{Switch}$($\Entry$ entry, int newSId)}{
    newSH = entry$\rightarrow$subhead)$\rightarrow$\newLocation\;
    \If{entry$\rightarrow$keyMin $\neq -\infty$}{
        leftEntry = registry.getByKey(keyMin)\;
        leftSH = leftEntry$\rightarrow$subhead\;
        \Do{$\neg$ response\label{line:switchSTEnd}}{
            \label{line:switchSTStart}
            \eIf{leftSH.id == me}{
                response = switchNextST( leftEntry$\rightarrow$subtail, newSH)\;
            } {
            leftSH = Send SwitchST(entry$\rightarrow$keyMin, newSH)\;
            }
            \If{leftSH == null}{
                response = true\;
            }
        }
    }
    entry$\rightarrow$subhead = newSH\;\label{line:switchRegistryStart}

    \For{$i \in$ serverList $-$ \{me, newSId\}} {
        response = Send SwitchServer( keyMax, newSH) to i\;
    }
    response = Send SwitchServer(keyMax, newSH) to newSId\;\label{line:switchRegistryEnd}
}
\Fn{\normalfont SwitchServerRecv($\Key$ keyMax, $\ItemRef$ newSH}{
    entry = registry.getByKey(keyMax)\;
    entry$\rightarrow$subhead = newSH\;
    return true\;
}
\Fn{\normalfont $\ItemRef$ SwitchSTRecv($\Key$ keyMin, $\ItemRef$ newSH)}{
    leftEntry = registry.getByKey(keyMin)\;
    leftSH = leftEntry$\rightarrow$subhead\;
    \If{leftSH.id == me}{
        \If{switchNextST( leftEntry$\rightarrow$subtail,newSH)}{
            return null\;
        }
        return leftSH$\rightarrow$\newLocation\;
    }
    return leftSH\;
}
\Fn{\normalfont bool switchNextST($\ItemRef$ leftST, $\ItemRef$ newSH)}{
    leftST$\rightarrow$\StartCount$\rightarrow$fetch\_add(1)\;
    \If{leftST$\rightarrow$\StartCount $<$ 0}{
        return false\;
    }
    leftST$\rightarrow$next = newSH\;
    leftST$\rightarrow$\EndCount$\rightarrow$fetch\_add(1)\;
    return true\;
}
\end{multicols}
\vspace*{-5mm}
\end{algorithm*}


\section{Correctness}
\label{sec:Correctness}
    We show the correctness of our algorithms by showing that the client operations are linearizable, and that they never get blocked by the background operations. We push the correctness of the background operations to Appendix \ref{sec:AppendixProofs} to save space. Here, however, we borrow the following observations proved in the appendix: \begin{enumerate*}
        \item There is only one active subhead to a sublist at any given time. 
        \item If an update on a sublist has checked that the sign of \StartCount is positive after incrementing it, then the \StartCount of the sublist cannot become negative until the update also increments its \EndCount. 
    \end{enumerate*}
    
\subsection{Linearizability}
    We show that the client operations are linearizable by identifying linearization points \cite{linearizability} (LP) for every possible output. DiLi is designed so that every operation can be said to take place in some specific machine entirely, by identifying the sublist that the operation takes place in. This is done by noting the currently active subhead for the sublist at the time when the operation is said to be linearized. Thus, our proof sketch for linearizability is by identifying the same linearization points as in the Harris List, while also showing that the same point in time was on the machine with the active sublist.

    \noindent
    \textbf{Find:} For a successful Find, the LP is immediately after the execution of the mark check returned false at Line \ref{line:searchMarkCheck} at the machine where Line \ref{line:SearchFoundReturn} also had its start count to be non-negative. For an unsuccessful Find, we have two cases -- \begin{enumerate*}
        \item If Line \ref{line:Search404Return} was ever evaluated as true, the LP is after the execution of the key comparison in that line, and it took place in the machine where the start count check on the same line returned false.
        \item If the above was not satisfied, then a subtail that upper bounds the key must have been reached by the check on Line \ref{Line:searchLastSubtailCheck}. Since a subtail does not get updated, instead of performing a subsequent start count check here, we instead linearize the find after the execution of Line \ref{line:searchStartCountNegativeCHeck} to be false and the operation took place in the machine where the latter was evaluated to be false.     
    \end{enumerate*}  

    \noindent
    \textbf{Remove:} For a successful Remove, the LP is immediately after the successful execution of CAS in Line \ref{Line:DeleteCAS}, where the next pointer of the item of the matching key gets marked. For an unsuccessful Remove, we have two cases -- \begin{enumerate*}
        \item If the search returned null, The LP is the same as an unsuccessful Find.
        \item If the search returned an item of a matching key, the LP is immediately after Line \ref{line:itemAlreadydeleted} returned true, i.e., when the node is already seen as marked/removed by another concurrent remove of the same key.
    \end{enumerate*}   

    \noindent
    \textbf{Insert:} For a successful Insert, the LP is immediately after the successful execution of CAS in Line \ref{line:InsertCAS}. For an unsuccessful Insert, we similarly have two cases in which the LP is the same as a successful Find.

\subsection{\DistributedLockFreedom of Client Operations}
\DistributedLockFreedom depends on the fact that any client operation utilizes at most four threads (three when there is no maintenance) and that the number of instructions executed by a delegating thread is bounded (proved in Appendix \ref{sec:appendixDelegationCountBound}).
When there is no background operation, being similar to a Harris List, a successful Remove causes at most two updates to a sublist -- one for marking (CAS in Line \ref{Line:DeleteCAS}) and one for delinking (CAS in Line \ref{line:delinkCAS}). A successful insert causes at most one update (CAS in Line \ref{line:InsertCAS}). The number of retries in Search is limited by the number of marked nodes to be delinked from concurrent Removes. Hence, no operation requires more retries than the number of concurrent updates \cite{harris2001pragmatic}.

During a Split operation, the search traversal (as denoted by the red lines in \Cref{alg:Main}) is only enhanced by an earlier truncation from the introduced subtail. An earlier truncation that occurs exactly at the item where a concurrent insert is taking place shortcuts to the next subhead (red lines of \Cref{alg:Main}, as the insertion is now expected to happen on the subhead instead of the item where the split is taking place, in order to preserve the key range property of the split sublists. By having zero additional search traversals for find and remove, and exactly one additional search traversal, the Harris List based operations that were lock-free when there is no split operation, are also lock-free during a Split operation.

During a Move operation (before Switch), the Find operation functions in the usual Harris List manner. The Remove and Insert operations additionally send Replication messages (replicates are asynchronous messages for which we do not wait for a response). Thus, Remove and Insert on a sublist only perform a constant amount of extra steps due to a concurrent Move on the same sublist.

During a Switch of a sublist, until the sublist entries of all servers are updated with the latest subhead, there will be an additional delegation to the right machine using the blue lines of \Cref{alg:Main} and \Cref{alg:Main} where the \StartCount values are checked to be negative. 
Thus, 
any concurrent client operation on the sublist
during this time
suffers from exactly one additional network hop and search traversal on the sublist. Since the search traversal and client operations were lock-free in the absence of a Switch, the client operations are lock-free even during the Switch.

\section{Empirical Evaluation} 
\label{sec:Empirical Evaluation}
    \subsection{Implementation}
    We implemented DiLi in C++20 in the Ubuntu 24.04.1 LTS operating system, utilized gRPC v1.72 \cite{gRPC} for client-server and ZeroMQ \cite{zmq} for server-server communications.
%
    Each machine that serves DiLi is assigned an initial key range to serve the list, chosen naively by a range partitioning on the key range of the list. As applicable to the experiment, dynamic re-partitioning is performed using $\Split$ and $\Move$, to improve performance. 
    The load balancer is a separate thread spawned in each machine that competes with threads performing the list operations (that also use the same set of server cores). It repeatedly traverses through all sublists held by the machine to split large sublists roughly in the middle, for quicker traversals.
    Using the same RPC framework, we also implemented the Java standard lock-free skip list \cite{FraserDissertation} in C++20. 
    
    We have also implemented a variation, where the linked list is packaged as a `library' that can be operated concurrently by multiple threads with no network communication overhead.
    The purpose of this experiment is to illustrate that while DiLi is designed to distribute a list across multiple machines, it provides a performance comparable to a skip list  when restricted to a single machine.
    

\subsection{Setup and Methodology}
    We perform four main experiments to evaluate the performance of DiLi:
    \begin{enumerate*}
        \item Multi-thread scalability of the DiLi library when compared to a lock-free skip list.
        \item A single-server multi-client comparison of DiLi with a lock-free skip list.
        \item Performance impact of delegations in a 2-machine setup (called the request locality experiment) and the benefit of an asynchronous move (called the move experiment) .
        \item A Distributed Scalability test to show that DiLi is scalable as we add more number of machines. 
    \end{enumerate*}
    
    The single-server and library experiments modify YCSB \cite{ycsb} workload template A to generate workloads of a Zipfian distribution. We load 1M keys into the list and then perform the experiment for a workload of 2M operations for five workload write proportions : 10\%, 25\%, 50\%, 75\% and 90\% write. The write proportion is equally split between inserts and removes, to keep the size of the list roughly the same. For the library experiment, we additionally evaluate five different loaded key sizes : 100K, 1M, 2M, 4M and 5M, while performing 2M operations of 50\% write workload.

    The experiments were performed using C7i instances from Amazon Web Services (AWS) as server machines for the list. The instance features an Intel(R) Xeon(R) Platinum 8488C (2.4GHz, 105MB L3 cache), with each core supporting 2 way hyper-threading and a $\sim12$ Gigabit network. The library and  single-server comparison utilized an 8 core instance of it with 32 GB of RAM, the 2-machine setups each had 4 cores and 16 GB of RAM, and the distributed list evaluation had multiple 2 core instances, each with 8 GB of RAM.

    For the distributed setup, the workload is additionally modified to demonstrate the effects of partitioning schemes. To obtain the maximum total throughput (number of operations completed per second) achieved in each setup, we perform several executions of the same experiment, varying the number of clients through a separate large compute instance, until a peak throughput is reached. Note here that the performance is evaluated on a low-latency network. Increased latency would give us a different set of results (much less throughput) due to the nature of the RPC framework and TCP connections.
    A dry run of 2 iterations is first performed before running either experiment to warm up the servers. Then each experiment is repeated 3 times and the average of those executions is used to compute the reported results.
    In the library setup, there is no network overhead that can overshadow the operation latencies, but they follow a similar pattern as the throughput. Hence, we skip the latency plots.
\begin{figure*}
\small{
    \begin{subfigure}[t]{0.32\textwidth}
        \centering
            \includegraphics[trim={1cm 0cm 2cm 0cm}, scale=0.32]{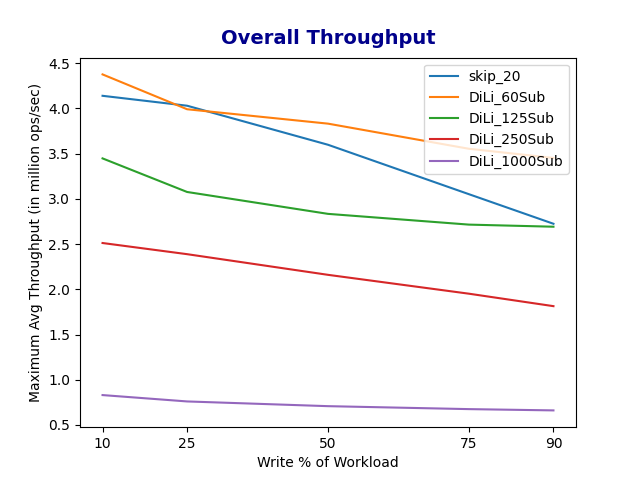}
            \caption{Maximum average throughput achieved when the write proportion is varied.}
            \label{fig:writeOverallThroughput}
    \end{subfigure}
    \begin{subfigure}[t]{0.32\textwidth}
        \centering
            \includegraphics[trim={1cm 0cm 2cm 0cm}, scale=0.32]{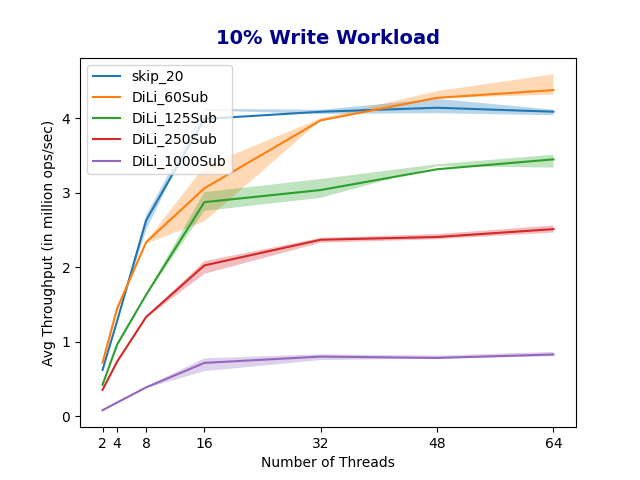}
            \caption{Scaling of skip list and DiLi for a read heavy workload.}
            \label{fig:write10Throughput}
    \end{subfigure}
    \begin{subfigure}[t]{0.32\textwidth}
        \centering
            \includegraphics[trim={1cm 0cm 2cm 0cm}, scale=0.32]{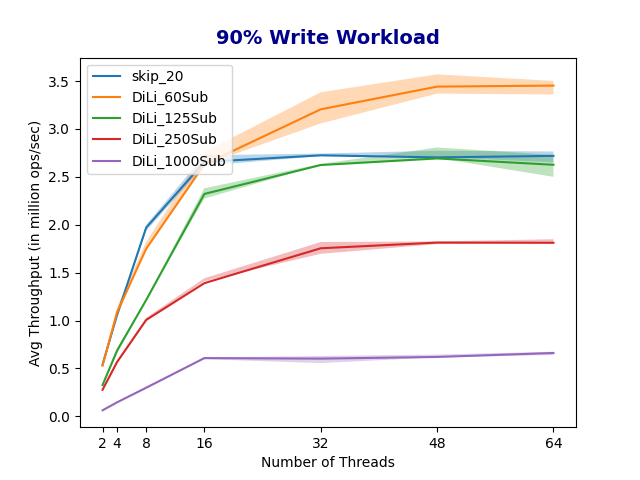}
            \caption{Scaling of skip list and DiLi for a write heavy workload.}
            \label{fig:write90Throughput}
    \end{subfigure}    
    \begin{subfigure}[t]{0.32\textwidth}
        \centering
            \includegraphics[trim={1cm 0cm 2cm 0cm}, scale=0.32]{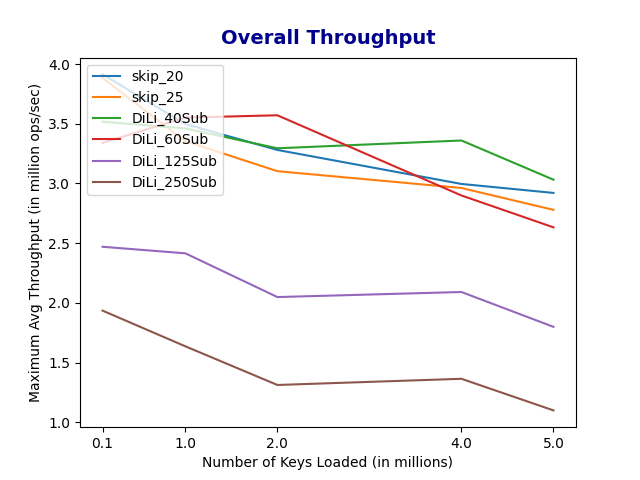}
            \caption{Maximum average throughput achieved when the key size is varied.
            }
            \label{fig:sizeOverallThroughput}
    \end{subfigure}
    \begin{subfigure}[t]{0.32\textwidth}
        \centering
            \includegraphics[trim={1cm 0cm 2cm 0cm}, scale=0.32]{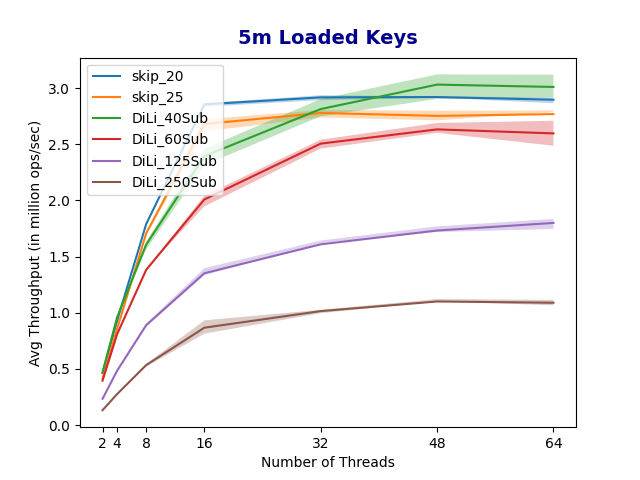}
            \caption{Scaling of skip list and DiLi for 5 million keys.}
            \label{fig:size5MThroughput}
    \end{subfigure}  
    \begin{subfigure}[t]{0.32\textwidth}
        \centering
            \includegraphics[trim={1cm 0cm 2cm 0cm}, scale=0.32]{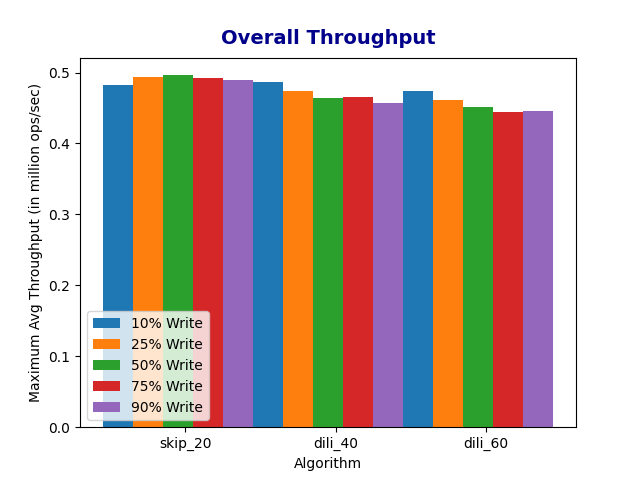}
            \caption{Single Machine Performance of DiLi as compared to skip lists.
            }
            \label{fig:singleMachineExp}
    \end{subfigure}
    \begin{subfigure}[t]{0.32\textwidth}
        \centering
            \includegraphics[trim={1cm 0cm 2cm 0cm}, scale=0.32]{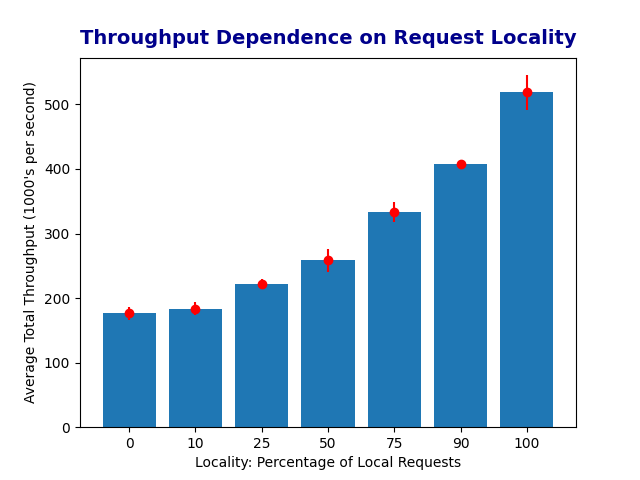}
            \caption{
            Throughput in a 2-machine setup depending on request locality.
            }
            \label{fig:localityExp}
    \end{subfigure}
    \begin{subfigure}[t]{0.32\textwidth}
        \centering
            \includegraphics[trim={1cm 0cm 2cm 0cm}, scale=0.32]{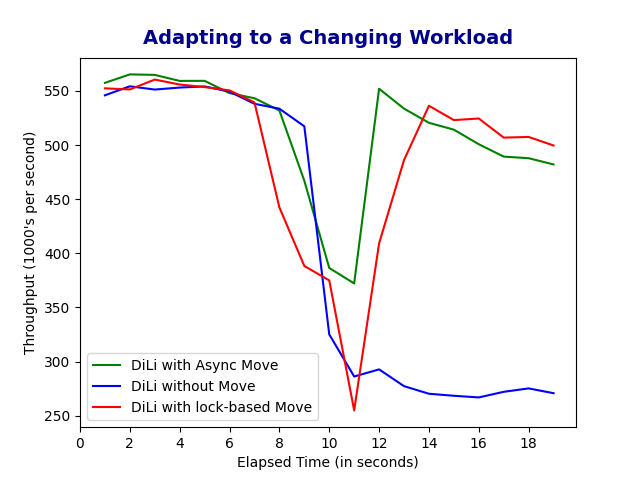}
            \caption{Benefit of an asynchronous $\Move$ operation for dynamic load balancing.
            }
            \label{fig:moveExp}
    \end{subfigure}
    \begin{subfigure}[t]{0.32\textwidth}
        \centering
            \includegraphics[trim={1cm 0cm 2cm 0cm}, scale=0.32]{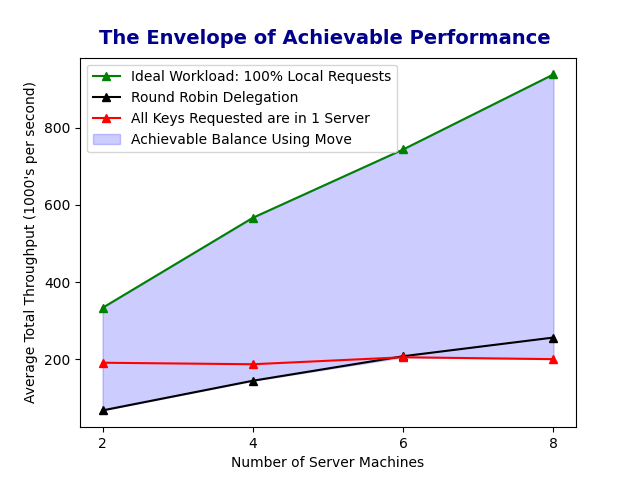}
            \caption{Distributed Scalability:
            The envelope of achievable performance with partitioning.}
            \label{fig:scalabilityExp}
    \end{subfigure}
    \label{fig:EmpiricalEvaluation}
    \caption{Performance of DiLi in library mode and distributed client-server architecture.}
}
\vspace*{-2mm}
\end{figure*}
\subsection{Results and Discussion}\label{subsec:resultsDiscussion}

\noindent\textbf{Library Experiments}: To evaluate DiLi as a multi-threaded library, we measure its throughput by varying the number of concurrent threads operating on it and compare it to Fraser's skip list. We perform this evaluation for \begin{enumerate*}
    \item varying write proportions;
    \item varying key sizes; 
\end{enumerate*} of the workloads.

\noindent\textbf{Write Proportion}: As we set the number of keys to 1 million, we use 20 as the optimal maximum level for skip lists. In \Cref{fig:writeOverallThroughput}, we observe that the performance of DiLi improves as we reduce the threshold size of a sublist. We run the $\Split$ method periodically on the same machine while the experiment is running to maintain this threshold size. Interestingly, $\Split$ terminates sufficiently fast to achieve this. When the threshold is set to 60, it even outperforms the skip list in read heavy workloads by 5.7\% and write heavy workloads by 26.7\%. As the write proportion increases, the performance of the skip list drops by about 34\%,  while DiLi$\_$60Sub (DiLi with 60 items per sublist threshold) drops only by 21\%. This shows that DiLi is also more stable than a skip list when write proportions are varied. Interestingly, as shown in \Cref{fig:write10Throughput} and \Cref{fig:write90Throughput}, the thread wise scaling of DiLi does not stop at 16 threads (= number of virtual cores),
This is important to note because the additional background thread for executing $\Split$ consumes shared CPU cores in varied amounts. We observe that DiLi outperforms the lock-free skip list\cite{FraserDissertation}, even when the background thread causes it to be oversubscribed for the same hardware.

\noindent\textbf{Key Size:} We also measure the library performance by varying the key size of the linked list, while keeping the write proportion at 50\%. In \Cref{fig:sizeOverallThroughput}, we observe that the performance of DiLi is only comparable to skip lists in high key sizes. DiLi$\_$40Sub only marginally outperforms skip list at 5M items by 3.8\% and DiLi$\_$60Sub outperforms skip list at 2M items by 8.847\%. We also note that there is no ``one size fits all" for DiLi. DiLi$\_$60Sub has 8\%  better throughput than DiLi$\_$40Sub at 2M, while being 13\% worse at 5M.

\noindent\textbf{Distributed Architecture Experiments}: Unlike a multi-threaded library, a distributed client-server model has the bottleneck of the network architecture. We compare single-server DiLi with its $\ItemRef$ overhead to a single-server skip list. Then we evaluated the distributed scalability of DiLi, which is our main contribution.

\noindent\textbf{Single-Server Setup}: In a client-server setup, we evaluate skip\_20 and DiLi in the presence of network overhead. We benchmarked the maximum throughput possible to be 979K using a \textit{ping} RPC. This reduces the possible throughput of any implementation to about a tenth of that of the library experiment. 
Unlike skip\_20, for DiLi, we even add the distribution overhead such as $\ItemRef$ pointer, checks for rerouting a request, keeping a port open to listen to other servers, etc. Interestingly, skip list and DiLi still have comparable throughput ($\sim$ 7\% difference) across workloads of all write proportions.

\noindent\textbf{2-Server Setup}: To introduce the challenges in distribution, we study the performance impact of request delegations  when a client request to a key is not local to the requested server by focusing on a 2-server setup. 

\noindent\textbf{Request Locality Experiment}: Here, we use two uniform distributions for the clients of each server to only perform requests based on their initial range partitioning. \Cref{fig:localityExp} shows that throughput improves by up to $\sim$  2.7 times when the locality of client requests is increased. This has a tremendous performance impact on dynamic workloads, and the $\Move$ operation can help maintain this client locality overtime.

\noindent\textbf{Move Experiment}: Here we initialize the list with our YCSB load of 1M keys, then have clients vary a 90\% Read workload overtime in two phases and plot throughput as time elapses (see \Cref{fig:moveExp}). In Phase 1 (approx. up to the 8th elapsed second), clients to the two servers (say S1 and S2) initially request only local keys. In Phase 2, clients of S2 request only a specific key range present in S1. By using $\Move$ on this key range asynchronously, we load balance requests to S2 to become local requests. Thus, DiLi can adapt to the changing workload and sustain the system throughput. For comparison, we also implemented a lock-based Move fine-grained on a sublist level. We observe that the lock-based move dips throughput $\sim$ 31\% more.

\noindent\textbf{Performance of Background Operations:} We additionally performed an experiment in a 2 server setup, where the first server has the entire key range for the list, and a second server starts without owning any part of the key range. We performed the 1M key load onto this setup and set the load balancer to transfer sublists to the idle server dynamically during an insert load, until both machines have a comparable number of items. During this test, we measured the latency of $\Split$ and $\Move$ (as depicted in Appendix \ref{subsec:appendixBackgroundOpTerminationResults}).  The average latency to Split was 2.65 ms and to Move a sublist it was 51.53 ms. Thus, we empirically verify that the asynchronous operations can terminate quickly even during an insert dominated workload. Note that DiLi only prioritizes the scalability of client operations and not that of the background operations.

\noindent\textbf{Distributed Scalability:} The true benefit of DiLi is scalability not only by multiple concurrent threads on a powerful machine, but by multiple weaker machines that serve the list. \Cref{fig:scalabilityExp} shows the throughput of DiLi with 1M loaded keys when 2, 4, 6 and 8 machines serve the list.
The green line demonstrates the maximum throughput (as obtained by 100\% request locality and even load distribution) with DiLi. The black line shows the performance with 0\% request locality but the key requests are evenly distributed to all machines in a round-robin manner. By letting Move transfer the keys to improve locality, we can increase performance up to the green line. One may wonder if this improvement can be achieved by changing the system model so that clients hold metadata to contact the correct machine every time. However, even in this case, a performance bottleneck can occur when all accessed keys are on the same server (say $S_0$) as depicted by the red line. In this case, $S_0$ can offload some of the keys to other servers using $\Move$ so that the overall throughput is closer to the green line. Thus, careful use of $\Move$ can provide various throughputs to DiLi, as depicted by the blue shaded region.
When requested keys are evenly distributed (e.g.,black and green lines), DiLi offers linear scaling of throughput. 

\subsection{Analysis from the Empirical Evaluation}
    We observe that DiLi can be used as a multi-threaded library that performs better than a skip list in write heavy workloads of key sizes as large as a million. DiLi is able to achieve this by having the overhead of a background thread on the same hardware setup. The sublist threshold size parameter can be adjusted based on the expected key size of the list to further improve the performance. Moreover, the resulting performance is maintained by only using one background thread for the setup. As we scale the setup to multiple machines, we can also choose to have multiple background threads per machine, and have them look at non-overlapping subsets of the registry for faster Splits as the key sizes get much larger.

    In a distributed setup, we observe that the network overhead throttles performance on top of the library implementations. One can reduce this overhead using RDMA networks. Regardless, because of dynamic load balancing, we are able to scale such performance with multiple machines. Note that we only used a simple load balancing logic implemented from our primitives. Tailoring a load balancing algorithm for various workloads can be a research on its own.
    
     In summary, DiLi provides performance comparable to skip lists when restricted to a single machine, where clients may be on the same machine or across a network. At the same time, DiLi can adapt to changing workload patterns by load balancing the list using its asynchronous $\Move$ operation. Finally, DiLi provides horizontal scalability by providing increased performance when more servers are added to maintain the list. 

\section{Related Work}
\label{sec:relatedWork}
    Linked lists are fundamental data structures. A sequential linked list is easy to implement. However, a concurrent linked list is challenging to design, and hence has been studied extensively for decades. 
Consequently, there exist many lock-free implementations -- 
\cite{Valois,harris2001pragmatic,Ruppert}; and a few wait-free implementations -- \cite{zhangWaitFreeLinkedList,Timnat}. 
The first lock-free implementation using the atomic CAS primitive (for a non-distributed list) is due to Valois \cite{Valois}. 
Harris \cite{harris2001pragmatic} gives a lock-free ordered list implementation. Among lock-free lists with no indexing and versioning, Harris List has been the state-of-the-art. Heller {\it et al.} \cite{optimisticlocklist} give a lock-based linked list design with a wait-free lookup operation. Our linear traversal of client operations resembles that of a Harris list. A wait-free ordered linked list implementation is given by Timnat {\it et al.} \cite{Timnat}.


Some papers focused on optimizing the cost of operations
in addition to correctness and progress guarantees established in the aforementioned papers.
An example is a lock-free list implementation due to Fomitchev and Ruppert \cite{Ruppert}, with worst-case linear amortized cost. 
\cite{harris2001pragmatic} and \cite{Ruppert} both
utilize a $\tt{HelpMarked}$ in their read operations.
They perform de-linking of marked nodes during any traversal. 
Zhang {\it et al.}\cite{zhangWaitFreeLinkedList} provided an unordered linked list where insertions always occur in the head position of the linked list.

Attiya and Hillel \cite{Attiya} give a concurrent implementation of a double-linked lock-free list using a double CAS operation. This list does not have a read operation. The same is implemented using only CAS by Sundell and Tsigas \cite{Tsigas}. 

The aforementioned implementations typically maintain the sequential entries in the linked list in different chunks of memory. To take advantage of entries that are instead on the same chunk,
Braginsky and Petrank \cite{chunkedlist} developed a concurrent linked list that maintains in each memory chunk, a certain number of entries within a defined minimum and maximum. Their approach shares some similarities with our key range component of a sublist. Such chunk mechanisms have also been explored in key-value stores that use a versioned linked list based search structure \cite{KiWi}. We note that this optimization is also applicable to the sublists of DiLi. We do not add this to our implementation to have a fair comparison with skip lists, which cannot have such an optimization across its levels.

The versioned linked list is a classic technique that arose from the need for read-only transactions in databases \cite{MultiVersioning}. Since we only support basic operations insert, remove, and find of a key in this paper, we lack the need for multi-versioning and hence do not compare our performance with versioned lists\cite{KiWi, jiffy}. Databases such as SingleStore\cite{SingleStore} also use lock-free skip lists to boost index search performance. These lock-free skip lists (\cite{sunderSkipList, FraserDissertation, fomitchenSkipList}), can be replaced with DiLi, as it offers comparable performance and multi-machine scalability. 



The aforementioned literature are all implementations on a single machine.  
Although there exist lock-based implementations of a distributed B-Tree in the literature \cite{minuet}, with regard to nonblocking distributed implementations on multiple machines, the literature is rare. The most related work is due to Abe and Yoshiday \cite{Distributeddoublylinkedlist} where a doubly linked distributed list is constructed. The presented strategy uses conflict detection and sequence numbers (with some assumptions) to break symmetries.  
It guarantees atomic insert and delete operations and non-blocking lookup operations, but the implementation is only obstruction-free. There also exists distributed data structures that unlike DiLi, are built for specialized hardware / operating systems. Alam et al. \cite{RangedQueryDistributedSkipList} developed a distributed skip list for systems that support FG-MPI\cite{FGMPI}.The skip list here is built for ranged queries, performs only one operation at a time, and has loose consistency guarantees for performance trade-offs using skip list shortcuts. Cell\cite{CellBTree} is a lock-based distributed B Tree that requires RDMA networks for communication.

To the best of our knowledge, DiLi is the first asynchronously distributable \distributedlockfree implementation of a linked list, and we believe it is an important step towards constructing other distributed, lock-free, linked data structures (trees, graphs, etc.). 


\section{Concluding Remarks}
\label{sec:Conclusion}
    In this paper, we developed DiLi, a distributed linked list data structure. The goal of this data structure is to preserve the client properties of Harris List (specifically, the lock-free nature) while providing horizontal scaling. Specifically, as the size of the data structure grows and as the number of operations performed on it in a given time increases, the data structure needs to be scaled to handle it. 

The distribution of the linked list created several challenges. These include ensuring that client operations continue to work correctly in a lock-free manner. Some properties of the list (e.g., linearizability, the number of elements in the list, the number of concurrent operations happening on the sublist) need to be preserved while the list is being split or moved. We developed an innovative approach through offsets of the sublist. This ensured that we can detect a virtual time when no operation is happening on the sublist that is being moved. 

Although designed as a linked list, the implementation of DiLi provides a performance comparable to skip lists \cite{FraserDissertation,sunderSkipList,fomitchenSkipList} on a single machine and even surpasses them as a library by up to 26\% throughput in write-heavy workloads. While these skip list implementations cannot provide horizontal scaling via distribution, DiLi is additionally capable of providing such scaling. 
Thus, DiLi provides a foundation for converting other high performance single machine data structures into distributed data structures. 
While the experiments illustrate a simple usage of the $\Split$ and $\Move$ primitives for YCSB workloads, future work can be designing load balancing strategies for any generic workload (such as TPC and synchrobench benchmarks).

The data structure considered in this paper, a linked list, differs from others, such as hash tables, in one important way. The elements in the hash table are not \textit{connected} to each other. In contrast, elements in a linked list are connected to each other by the notion of the \textit{next element}. These relations are preserved by the split and move operations in DiLi. It is therefore possible to ensure that these operations can be added to DiLi while ensuring that their lock-free nature is preserved even while the list is being split or moved. This will open up the possibility of horizontally scaling other lock-free data structures (e.g., trees, graphs) while preserving their lock-free nature. Another future direction would be to introduce operations that span multiple machines and exploit the linked list structure such as ranged queries and snapshots. We believe that this paper will be foundational for such future work.

Finally, this paper focused on the scalability aspect of distribution while maintaining lock-freedom of client operations. Another important direction would be fault tolerance, such as additional maintenance of the list data in a replication fashion.



\section*{Acknowledgment}
 This work was supported in part through computational resources and services provided by the Institute for Cyber-Enabled Research, Cloud Computing Fellowship Program at Michigan State University. The authors also thank Eliezer Amponsah for his invaluable contributions to the framework built for the empirical evaluation.
{
\bibliographystyle{plain}
\bibliography{ref}

@InProceedings{zhangWaitFreeLinkedList,
author="Zhang, Kunlong
and Zhao, Yujiao
and Yang, Yajun
and Liu, Yujie
and Spear, Michael",
title="Practical Non-blocking Unordered Lists",
booktitle="DISC",
year="2013",
pages="239--253",
isbn="978-3-642-41527-2"
}

@ARTICLE{hazardPointers,
  author={Michael, M.M.},
  journal={IEEE Transactions on Parallel and Distributed Systems}, 
  title={Hazard pointers: safe memory reclamation for lock-free objects}, 
  year={2004},
  volume={15},
  number={6},
  pages={491-504},
  doi={10.1109/TPDS.2004.8}}

@inproceedings{harris2001pragmatic,
  title={A pragmatic implementation of non-blocking linked-lists},
  author={Harris, Timothy L},
  booktitle={DISC},
  pages={300--314},
  year={2001},
  organization={Springer}
}

@InProceedings{MultiCAS,
author="Harris, Timothy L.
and Fraser, Keir
and Pratt, Ian A.",
title="A Practical Multi-word Compare-and-Swap Operation",
booktitle="DISC",
year="2002",
pages="265--279",
isbn="978-3-540-36108-4"
}

@inproceedings{Distributeddoublylinkedlist,
  author       = {Kota Abe and
                  Mikio Yoshiday},

  title        = {Constructing distributed doubly linked lists without distributed locking},
  booktitle    = {P2P},
  pages        = {1--10},
  publisher    = {{IEEE}},
  year         = {2015},
  url          = {https://doi.org/10.1109/P2P.2015.7328521},
  doi          = {10.1109/P2P.2015.7328521},
  bibsource    = {dblp computer science bibliography, https://dblp.org}
}

@InProceedings{optimisticlocklist,
author="Heller, Steve
and Herlihy, Maurice
and Luchangco, Victor
and Moir, Mark
and Scherer, William N.
and Shavit, Nir",
title="A Lazy Concurrent List-Based Set Algorithm",
booktitle="OPODIS",
year="2006",
pages="3--16",
isbn="978-3-540-36322-4"
}

@article{linearizability,
author = {Herlihy, Maurice P. and Wing, Jeannette M.},
title = {Linearizability: A Correctness Condition for Concurrent Objects},
year = {1990},
issue_date = {July 1990},
publisher = {Association for Computing Machinery},
address = {New York, NY, USA},
volume = {12},
number = {3},
issn = {0164-0925},
url = {https://doi.org/10.1145/78969.78972},
doi = {10.1145/78969.78972},
journal = {ACM Trans. Program. Lang. Syst.},
month = {jul},
pages = {463–492},
numpages = {30}
}

@article{Tsigas,
  author       = {H{\aa}kan Sundell and
                  Philippas Tsigas},
  title        = {Lock-free deques and doubly linked lists},
  journal      = {J. Parallel Distributed Comput.},
  volume       = {68},
  number       = {7},
  pages        = {1008--1020},
  year         = {2008},
  url          = {https://doi.org/10.1016/j.jpdc.2008.03.001},
  doi          = {10.1016/J.JPDC.2008.03.001},
  bibsource    = {dblp computer science bibliography, https://dblp.org}
}

@inproceedings{Timnat,
  author       = {Shahar Timnat and
                  Anastasia Braginsky and
                  Alex Kogan and
                  Erez Petrank},
  title        = {Wait-Free Linked-Lists},
  booktitle    = {OPODIS},
  pages        = {330--344},
  publisher    = {Springer},
  year         = {2012},
  url          = {https://doi.org/10.1007/978-3-642-35476-2\_23},
  doi          = {10.1007/978-3-642-35476-2\_23},
  bibsource    = {dblp computer science bibliography, https://dblp.org}
}

@inproceedings{Ruppert,
  author       = {Mikhail Fomitchev and
                  Eric Ruppert},
  editor       = {Soma Chaudhuri and
                  Shay Kutten},
  title        = {Lock-free linked lists and skip lists},
  booktitle    = {PODC},
  pages        = {50--59},
  publisher    = {{ACM}},
  year         = {2004},
  url          = {https://doi.org/10.1145/1011767.1011776},
  doi          = {10.1145/1011767.1011776},
  bibsource    = {dblp computer science bibliography, https://dblp.org}
}

@inproceedings{Attiya,
  author       = {Hagit Attiya and
                  Eshcar Hillel},
  editor       = {Shlomi Dolev},
  title        = {Built-In Coloring for Highly-Concurrent Doubly-Linked Lists},
  booktitle    = {DISC},
  pages        = {31--45},
  publisher    = {Springer},
  year         = {2006},
  url          = {https://doi.org/10.1007/11864219\_3},
  doi          = {10.1007/11864219\_3},
  bibsource    = {dblp computer science bibliography, https://dblp.org}
}

@inproceedings{Valois,
  author       = {John D. Valois},
  editor       = {James H. Anderson},
  title        = {Lock-Free Linked Lists Using Compare-and-Swap},
  booktitle    = {PODC},
  pages        = {214--222},
  publisher    = {{ACM}},
  year         = {1995},
}

@inproceedings{chunkedlist,
  author       = {Anastasia Braginsky and
                  Erez Petrank},
  editor       = {Marcos Kawazoe Aguilera and
                  Haifeng Yu and
                  Nitin H. Vaidya and
                  Vikram Srinivasan and
                  Romit Roy Choudhury},
  title        = {Locality-Conscious Lock-Free Linked Lists},
  booktitle    = {ICDCN},
  pages        = {107--118},
  publisher    = {Springer},
  year         = {2011},
}

@article{
PostMooresLaw,
author = {Charles E. Leiserson  and Neil C. Thompson  and Joel S. Emer  and Bradley C. Kuszmaul  and Butler W. Lampson  and Daniel Sanchez  and Tao B. Schardl },
title = {There’s plenty of room at the Top: What will drive computer performance after Moore’s law?},
journal = {Science},
volume = {368},
number = {6495},
pages = {eaam9744},
year = {2020},
doi = {10.1126/science.aam9744},
URL = {https://www.science.org/doi/abs/10.1126/science.aam9744},
eprint = {https://www.science.org/doi/pdf/10.1126/science.aam9744}}

@article{KiWi,
author = {Basin, Dmitry and Bortnikov, Edward and Braginsky, Anastasia and Golan-Gueta, Guy and Hillel, Eshcar and Keidar, Idit and Sulamy, Moshe},
title = {KiWi: A Key-value Map for Scalable Real-time Analytics},
year = {2020},
issue_date = {September 2020},
publisher = {Association for Computing Machinery},
address = {New York, NY, USA},
volume = {7},
number = {3},
issn = {2329-4949},
url = {https://doi.org/10.1145/3399718},
doi = {10.1145/3399718},
journal = {ACM Trans. Parallel Comput.},
month = jun,
articleno = {16},
numpages = {28}
}

@article{minuet,
author = {Sowell, Benjamin and Golab, Wojciech and Shah, Mehul A.},
title = {Minuet: a scalable distributed multiversion B-tree},
year = {2012},
issue_date = {May 2012},
publisher = {VLDB Endowment},
volume = {5},
number = {9},
issn = {2150-8097},
url = {https://doi.org/10.14778/2311906.2311915},
doi = {10.14778/2311906.2311915},
journal = {Proc. VLDB Endow.},
month = may,
pages = {884–895},
numpages = {12}
}

@Inbook{MultiVersioning,
author="Keidar, Idit
and Perelman, Dmitri",
title="Multi-versioning in Transactional Memory",
bookTitle="Transactional Memory. Foundations, Algorithms, Tools, and Applications: COST Action Euro-TM IC1001",
year="2015",
publisher="Springer International Publishing",
address="Cham",
pages="150--165",
isbn="978-3-319-14720-8",
doi="10.1007/978-3-319-14720-8_7",
url="https://doi.org/10.1007/978-3-319-14720-8_7"
}

@TechReport{FraserDissertation,
  author =	 {Fraser, Keir},
  title = 	 {{Practical lock-freedom}},
  year = 	 2004,
  month = 	 feb,
  url = 	 {https://www.cl.cam.ac.uk/techreports/UCAM-CL-TR-579.pdf},
  institution =  {University of Cambridge, Computer Laboratory},
  doi = 	 {10.48456/tr-579},
  number = 	 {UCAM-CL-TR-579}
}

@inproceedings{sunderSkipList,
author = {Sundell, H\r{a}kan and Tsigas, Philippas},
title = {Scalable and lock-free concurrent dictionaries},
year = {2004},
isbn = {1581138121},
publisher = {Association for Computing Machinery},
address = {New York, NY, USA},
url = {https://doi.org/10.1145/967900.968188},
doi = {10.1145/967900.968188},
booktitle = {Proceedings of the 2004 ACM Symposium on Applied Computing},
pages = {1438–1445},
numpages = {8},
location = {Nicosia, Cyprus},
series = {SAC '04}
}

@inproceedings{fomitchenSkipList,
author = {Fomitchev, Mikhail and Ruppert, Eric},
title = {Lock-free linked lists and skip lists},
year = {2004},
isbn = {1581138024},
publisher = {Association for Computing Machinery},
address = {New York, NY, USA},
url = {https://doi.org/10.1145/1011767.1011776},
doi = {10.1145/1011767.1011776},
booktitle = {Proceedings of the Twenty-Third Annual ACM Symposium on Principles of Distributed Computing},
pages = {50–59},
numpages = {10},
location = {St. John's, Newfoundland, Canada},
series = {PODC '04}
}

@inproceedings{jiffy,
author = {Kobus, Tadeusz and Kokoci\'{n}ski, Maciej and Wojciechowski, Pawe\l{} T.},
title = {Jiffy: a lock-free skip list with batch updates and snapshots},
year = {2022},
isbn = {9781450392044},
publisher = {Association for Computing Machinery},
address = {New York, NY, USA},
url = {https://doi.org/10.1145/3503221.3508437},
doi = {10.1145/3503221.3508437},
booktitle = {Proceedings of the 27th ACM SIGPLAN Symposium on Principles and Practice of Parallel Programming},
pages = {400–415},
numpages = {16},
location = {Seoul, Republic of Korea},
series = {PPoPP '22}
}

@book{herlihybook,
author = {Herlihy, Maurice and Shavit, Nir},
title = {The Art of Multiprocessor Programming, Revised Reprint},
year = {2012},
isbn = {9780123973375},
publisher = {Morgan Kaufmann Publishers Inc.},
address = {San Francisco, CA, USA},
edition = {1st}
}

@inproceedings{CAP,
author = {Brewer, Eric A.},
title = {Towards robust distributed systems (abstract)},
year = {2000},
isbn = {1581131836},
publisher = {Association for Computing Machinery},
address = {New York, NY, USA},
url = {https://doi.org/10.1145/343477.343502},
doi = {10.1145/343477.343502},
booktitle = {Proceedings of the Nineteenth Annual ACM Symposium on Principles of Distributed Computing},
pages = {7},
location = {Portland, Oregon, USA},
series = {PODC '00}
}

@misc{gRPC, 
key = {Google},
title = {{gRPC}},
year = {2016},
month = aug,
howpublished = {\url{https://grpc.io/}},
}

@misc{hbase,
key = {Apache},
title = {{Apache HBase - A distributed, scalable, big data store.}},
year = {2008},
howpublished = {\url{https://hbase.apache.org/}}
}

@misc{leveldb,
key = {Google},
title = {{LevelDB - A fast key-value storage library.}},
year = {2011},
howpublished = {\url{https://github.com/google/leveldb}}
}

@misc{rocksdb,
key = {Meta},
title = {{RocksDB: A Persistent Key-Value Store for Flash and RAM Storage.}},
year = {2012},
howpublished = {\url{https://github.com/facebook/rocksdb}}
}

@inproceedings{ycsb,
author = {Cooper, Brian F. and Silberstein, Adam and Tam, Erwin and Ramakrishnan, Raghu and Sears, Russell},
title = {Benchmarking cloud serving systems with YCSB},
year = {2010},
isbn = {9781450300360},
publisher = {Association for Computing Machinery},
address = {New York, NY, USA},
url = {https://doi.org/10.1145/1807128.1807152},
doi = {10.1145/1807128.1807152},
booktitle = {Proceedings of the 1st ACM Symposium on Cloud Computing},
pages = {143–154},
numpages = {12},
location = {Indianapolis, Indiana, USA},
series = {SoCC '10}
}

@misc{intel5LevelPaging,
	author = {},
	title = {5-{L}evel {P}aging and 5-{L}evel {E}{P}{T} {W}hite {P}aper --- intel.com},
	howpublished = {\url{https://www.intel.com/content/www/us/en/content-details/671442/5-level-paging-and-5-level-ept-white-paper.html}},
	year = {},
	note = {[Accessed 19-08-2025]},
}

@inproceedings{SingleStore,
author = {Prout, Adam and Wang, Szu-Po and Victor, Joseph and Sun, Zhou and Li, Yongzhu and Chen, Jack and Bergeron, Evan and Hanson, Eric and Walzer, Robert and Gomes, Rodrigo and Shamgunov, Nikita},
title = {Cloud-Native Transactions and Analytics in SingleStore},
year = {2022},
isbn = {9781450392495},
publisher = {Association for Computing Machinery},
address = {New York, NY, USA},
url = {https://doi.org/10.1145/3514221.3526055},
doi = {10.1145/3514221.3526055},
booktitle = {Proceedings of the 2022 International Conference on Management of Data},
pages = {2340–2352},
numpages = {13},
location = {Philadelphia, PA, USA},
series = {SIGMOD '22}
}

@inproceedings{RangedQueryDistributedSkipList,
author = {Alam, Sarwar and Kamal, Humaira and Wagner, Alan},
title = {A scalable distributed skip list for range queries},
year = {2014},
isbn = {9781450327497},
publisher = {Association for Computing Machinery},
address = {New York, NY, USA},
url = {https://doi.org/10.1145/2600212.2600712},
doi = {10.1145/2600212.2600712},
booktitle = {Proceedings of the 23rd International Symposium on High-Performance Parallel and Distributed Computing},
pages = {315–318},
numpages = {4},
location = {Vancouver, BC, Canada},
series = {HPDC '14}
}

@inproceedings{CellBTree,
author = {Mitchell, Christopher and Montgomery, Kate and Nelson, Lamont and Sen, Siddhartha and Li, Jinyang},
title = {Balancing CPU and network in the cell distributed B-tree store},
year = {2016},
isbn = {9781931971300},
publisher = {USENIX Association},
address = {USA},
booktitle = {Proceedings of the 2016 USENIX Conference on Usenix Annual Technical Conference},
pages = {451–464},
numpages = {14},
location = {Denver, CO, USA},
series = {USENIX ATC '16}
}

@INPROCEEDINGS{FGMPI,
  author={Kamal, Humaira and Wagner, Alan},
  booktitle={2010 IEEE International Symposium on Parallel \& Distributed Processing, Workshops and Phd Forum (IPDPSW)}, 
  title={FG-MPI: Fine-grain MPI for multicore and clusters}, 
  year={2010},
  volume={},
  number={},
  pages={1-8},
  doi={10.1109/IPDPSW.2010.5470773}}

@article{synchrobench,
author = {Gramoli, Vincent},
title = {More than you ever wanted to know about synchronization: synchrobench, measuring the impact of the synchronization on concurrent algorithms},
year = {2015},
issue_date = {August 2015},
publisher = {Association for Computing Machinery},
address = {New York, NY, USA},
volume = {50},
number = {8},
issn = {0362-1340},
url = {https://doi.org/10.1145/2858788.2688501},
doi = {10.1145/2858788.2688501},
journal = {SIGPLAN Not.},
month = jan,
pages = {1–10},
numpages = {10}
}

@book{zmq,
author = {Akgul, Faruk},
title = {ZeroMQ},
year = {2013},
isbn = {178216104X},
publisher = {Packt Publishing}
}

@inproceedings{shardingInBlockChains,
author = {Dang, Hung and Dinh, Tien Tuan Anh and Loghin, Dumitrel and Chang, Ee-Chien and Lin, Qian and Ooi, Beng Chin},
title = {Towards Scaling Blockchain Systems via Sharding},
year = {2019},
isbn = {9781450356435},
publisher = {Association for Computing Machinery},
address = {New York, NY, USA},
url = {https://doi.org/10.1145/3299869.3319889},
doi = {10.1145/3299869.3319889},
booktitle = {Proceedings of the 2019 International Conference on Management of Data},
pages = {123–140},
numpages = {18},
location = {Amsterdam, Netherlands},
series = {SIGMOD '19}
}

@inproceedings{shardingInBlockchains2,
author = {Zamani, Mahdi and Movahedi, Mahnush and Raykova, Mariana},
title = {RapidChain: Scaling Blockchain via Full Sharding},
year = {2018},
isbn = {9781450356930},
publisher = {Association for Computing Machinery},
address = {New York, NY, USA},
url = {https://doi.org/10.1145/3243734.3243853},
doi = {10.1145/3243734.3243853},
booktitle = {Proceedings of the 2018 ACM SIGSAC Conference on Computer and Communications Security},
pages = {931–948},
numpages = {18},
location = {Toronto, Canada},
series = {CCS '18}
}

@inproceedings{ShardingInBlockchains3,
author = {Wang, Gang and Shi, Zhijie Jerry and Nixon, Mark and Han, Song},
title = {SoK: Sharding on Blockchain},
year = {2019},
isbn = {9781450367325},
publisher = {Association for Computing Machinery},
address = {New York, NY, USA},
url = {https://doi.org/10.1145/3318041.3355457},
doi = {10.1145/3318041.3355457},
booktitle = {Proceedings of the 1st ACM Conference on Advances in Financial Technologies},
pages = {41–61},
numpages = {21},
location = {Zurich, Switzerland},
series = {AFT '19}
}

@InProceedings{GraphDBSurvey,
author="Lopez-Veyna, Jaime I.
and Castillo-Zu{\~{n}}iga, Ivan
and Ortiz-Garcia, Mariana",
editor="Mejia, Jezreel
and Mu{\~{n}}oz, Mirna
and Rocha, {\'A}lvaro
and Hern{\'a}ndez-Nava, V{\'i}ctor",
title="A Review of Graph Databases",
booktitle="New Perspectives in Software Engineering",
year="2023",
publisher="Springer International Publishing",
address="Cham",
pages="180--195",
isbn="978-3-031-20322-0"
}
}

\appendices
    \section{Registry Operations}\label{sec:appendixRegistry}

    \begin{algorithm}
    \vspace*{-7mm}
    \scriptsize
    \caption{Registry Operations}
    \SetKwProg{Fn}{}{:}{end}
    \label{alg:Registry}
    \begin{multicols}{2}

    $\Registry$ *registry\;
    \Fn{\normalfont \text{void} addEntry($\Entry$ *entry)}{
    $\Key$ newKey = entry$\rightarrow$keyMin\;
    \Do{$\neg$CAS(registry, currReg, newReg)}{
        currReg = registry\;
        Registry* newReg\;
        \For{$i=0$; (i $<$ currReg$\rightarrow$ size $\wedge$ currReg $\rightarrow$ entries[i]$\rightarrow$ keyMin $<$ newKey ); i = i + 1 } {
            newReg$\rightarrow$entries[i] = currReg$\rightarrow$entries[i]\;
        }
        newReg$\rightarrow$entries[i] = entry\;
        \For{(; i $<$ currReg$\rightarrow$ size; i = i + 1)} {
        newReg$\rightarrow$entries[i+1] = currReg$\rightarrow$entries[i]\;
        }
        i = i + 1\;
        newReg$\rightarrow$entries[i] = null\;
        newReg$\rightarrow$size = i\;
    }
}

\Fn{\normalfont $\Entry$ *getByKey($\Key$ key)}{
    currReg $=$ registry\;
    $\text{entries} = \text{currReg}\rightarrow  \text{entries}$\;
    
    left $=$ 0, right $=$ currReg$\rightarrow$ size - 1\; 
    \While{left $\leq$ right}{
     middle $=$ (left $+$ right)$/$2\;
     \eIf{key $\leq$ entries[middle]$\rightarrow$keyMin}{
        right $=$ middle$-$1\;     
     } {
        \eIf{
        key $\leq$ entries[middle]$\rightarrow$keyMax
        }{
            left $=$ middle $+$1\;
        }
        {
            return entries[middle]\;
        }
     }
    }
    return null\;
    
}
\end{multicols}
\vspace*{-5mm}
\end{algorithm}

    As depicted in \Cref{alg:Registry}, the Registry features two main operations -- getByKey(key) and addEntry(entry). AddEntry copies the registry to a new registry, while placing the new entry at the right position of the array, which is kept sorted by the keyMin of the entries. In our implementation, we additionally use \cite{hazardPointers} to safely reclaim the old registry (memory stored in the currReg pointer variable in the addEntry() routine) after the last pointer to it goes out of scope. The operation getByKey() reads the value of the current registry pointer and does the classic binary search on the registry as an array. Note that since we allow multi read (used by client operations), single write (used only by Split) for the registry, any implementation of a lock-based binary search tree that has a wait free search could be used instead of our simple illustration with arrays and copy-on-writes. A removeEntry() method can be implemented using a similar copy-on-write technique.

\newpage
\section{Merge Operation}
\label{sec:appendixMerge}

    The algorithm for the Merge operation is shown in \Cref{alg:Merge}. The operation takes the two registry entries of the sublists that are in the same server and next to each other as input, then merges the right sublist with the left sublist, and returns the left sublist entry as the response. The algorithm is very similar to Split, but the logic occurs in a different order. First, The left sublist entry is updated with the merged key range. Then the right sublist entry is removed from the index. This allows subsequent traversals of the client operations to be possible just from the left sublist. Then \StartCount and \EndCount of the items in the right sublist are updated with the left sublist counters.
    
    Now all that is left before announcing the merge is to remove the subtail and subhead present in the middle of the merged sublist. This is done using a restricted double compare single swap (RDCSS) that is implemented using 3 CAS operations \cite{MultiCAS}. The operation takes two pairs of $<$variable, value$>$ pairs (hence the first 4 parameters) for comparison and replaces the first variable with the fifth parameter. For the merge to remove the subtail subhead block, the operation repeatedly tries to update the next pointer of the last element of the left sublist to the first element of the right sublist. To make it asynchronous, it must succeed only when no insertions take place on the subhead to be removed. Hence, a double comparison is performed for a single swap. After this, the new offset is computed in the same way as $\Split$ (\Cref{subsec:split}, by finding the difference between the counters during a momentary write-free period. The updated index is then announced to the other machines via RegisterMergedSublist message. 

\begin{algorithm}
\scriptsize
\caption{Merge Operation}
\label{alg:Merge}
\Fn{\normalfont $\Entry$ *Merge($\Entry$* leftEntry, $\Entry$ rightEntry)} {

    leftEntry$\rightarrow$subtail$\rightarrow$ keyMax = leftEntry$\rightarrow$ keyMin\; 
    leftEntry$\rightarrow$ keyMax = rightEntry$\rightarrow$ keyMax\;
    leftEntry$\rightarrow$subtail = rightEntry$\rightarrow$subtail\;

    registry.removeEntry(rightEntry)\;

    curr = rightEntry$\rightarrow$SH\;
    \Do{prev$\rightarrow$key $\neq$ ST\_KEY}{
        prev = curr\;
        curr$\rightarrow$\StartCount = leftEntry$\rightarrow$subhead$\rightarrow$\StartCount\;
        curr$\rightarrow$\EndCount = leftEntry$\rightarrow$subhead$\rightarrow$\EndCount\;
        curr = curr$\rightarrow$next\;
    }
    
    \Do{$\neg$ RDCSS(leftLast$\rightarrow$next, rightEntry$\rightarrow$ subtail, rightFirst$\rightarrow$next, rightFirstNext, rightFirst)}{
        leftLast = leftEntry$\rightarrow$subhead\;
        \While{leftLast$\rightarrow$next$\rightarrow$key $\neq$ ST\_KEY}{
            leftLast = leftLast$\rightarrow$next\;
        }
        rightFirst = rightEntry$\rightarrow$subhead$\rightarrow$next\;
        rightFirstNext = rightFirst$\rightarrow$next\;
    }

    \Do{($a_1$ + $a_2$) $\neq$ (leftEntry$\rightarrow$\offset + rightEntry$\rightarrow$\offset) }{
    $a_1$ = leftEntry$\rightarrow$\StartCount - leftEntry$\rightarrow$\EndCount\;
    $a_2$ = rightEntry$\rightarrow$\StartCount - rightEntry$\rightarrow$\EndCount\;
    }

    leftEntry$\rightarrow$\offset = $a_1$\;

    \For{$i \in $(\normalfont{serverList} $-$ \{\normalfont{me}\})}{
        response = Send RegisterMergedSublist(rightEntry$\rightarrow$keyMax)\;
    }
    
    return leftEntry\;
}

\Fn{\normalfont bool RegisterMergedSublistRecv(Key keyMid, $\ItemRef$ SH)} {
    $\Entry$ rightEntry = registry.getByKey(keyMid)\;
    $\Entry$ leftEntry = registry.getByKey(keyMid - 1)\;
    leftEntry$\rightarrow$keyMax = keyMid\;
    registry.removeEntry(rightEntry)\;
    return true\;
}
\vspace*{-3mm}
\end{algorithm}

\section{Supplementary Results}
\label{sec:AppendixResults}

\subsubsection{Thread-Wise Variation of Various Write Workloads}
\label{subsec:appendixLibraryExperimentExtendedResults}
\begin{figure*}
\small{
    \begin{subfigure}[t]{0.32\textwidth}
        \centering
            \includegraphics[trim={1cm 0cm 2cm 0cm}, scale=0.32]{Figures/library_write_workloads/Write_10_Max_Avg_Th.png}
            \caption{Maximum average throughput achieved when write proportion is 10\%.}
            \label{fig:appWrite10}
    \end{subfigure}
    \begin{subfigure}[t]{0.32\textwidth}
        \centering
            \includegraphics[trim={1cm 0cm 2cm 0cm}, scale=0.32]{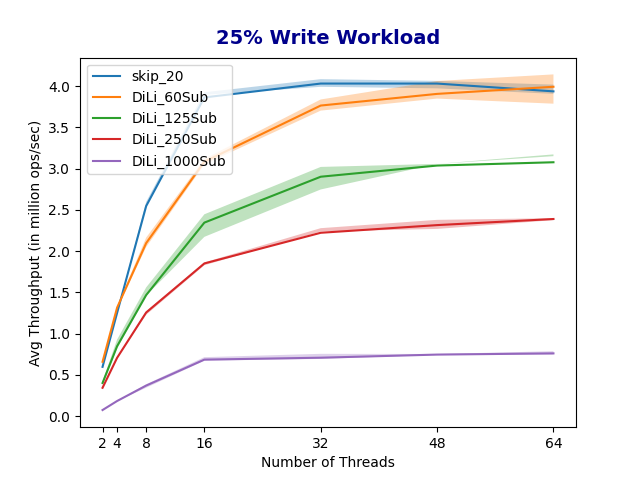}
            \caption{Maximum average throughput achieved when write proportion is 25\%. }
            \label{fig:appWrite25}
    \end{subfigure}
    \begin{subfigure}[t]{0.32\textwidth}
        \centering
            \includegraphics[trim={1cm 0cm 2cm 0cm}, scale=0.32]{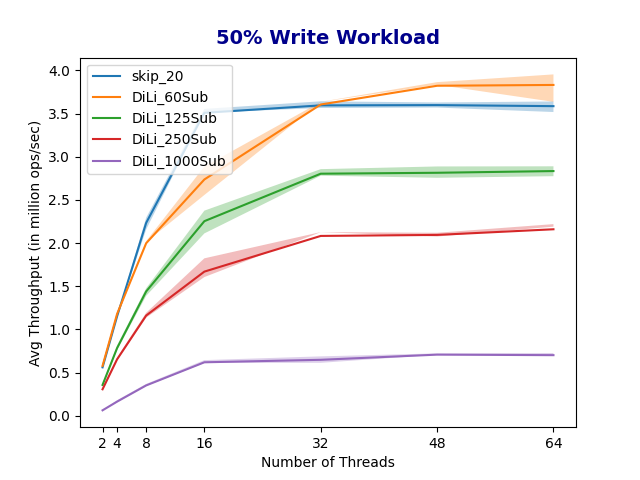}
            \caption{Maximum average throughput achieved when write proportion is 50\%.}
            \label{fig:appWrite50}
    \end{subfigure}
    \begin{subfigure}[t]{0.32\textwidth}
        \centering
            \includegraphics[trim={1cm 0cm 2cm 0cm}, scale=0.32]{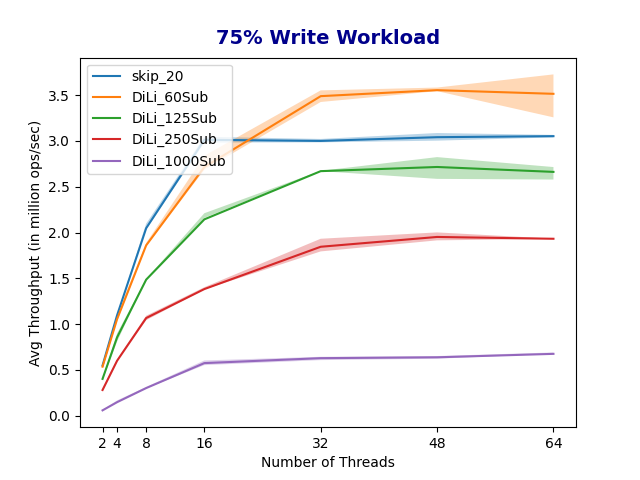}
            \caption{Maximum average throughput achieved when write proportion is 75\%.}
            \label{fig:appWrite75}
    \end{subfigure}   
    \begin{subfigure}[t]{0.32\textwidth}
        \centering
            \includegraphics[trim={1cm 0cm 2cm 0cm}, scale=0.32]{Figures/library_write_workloads/Write_90_Max_Avg_Th.png}
            \caption{Maximum average throughput achieved when write proportion is 90\%.}
            \label{fig:appWrite90}
    \end{subfigure}   }
    \caption{Performance of DiLi and Skip List when used as a library to execute workloads of various write proportions. Key size is maintained to be 1 million throughout.}
    \label{fig:AppWriteWorkloadPerformance}
\end{figure*}

    \Cref{fig:AppWriteWorkloadPerformance} shows the thread-wise results for all write workloads that we have experimented.

\subsubsection{Thread-Wise Variation for Different Sizes of List}
\label{subsec:appendixSizeExperimentExtendedResults}
\begin{figure*}
\small {
    \begin{subfigure}[t]{0.32\textwidth}
        \centering
            \includegraphics[trim={1cm 0cm 2cm 0cm}, scale=0.32]{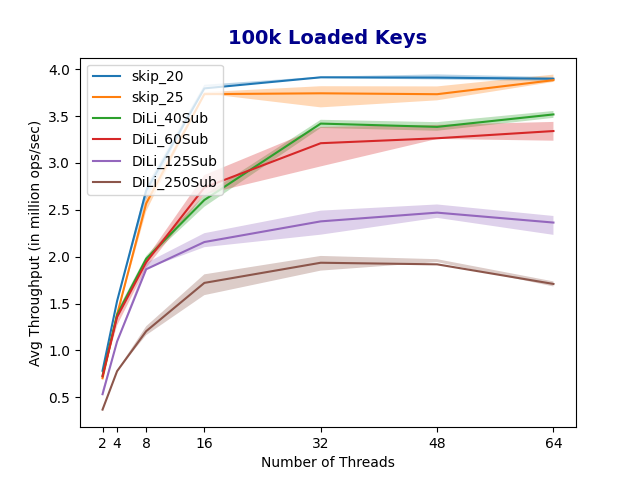}
            \caption{Maximum size throughput achieved when size of the list is 100k.}
            \label{fig:appSize100k}
    \end{subfigure}
    \begin{subfigure}[t]{0.32\textwidth}
        \centering
            \includegraphics[trim={1cm 0cm 2cm 0cm}, scale=0.32]{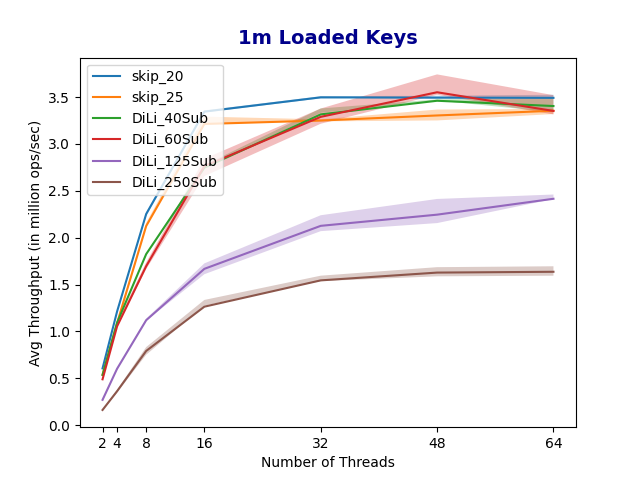}
            \caption{Maximum size throughput achieved when size of the list is 1m. }
            \label{fig:appSize1m}
    \end{subfigure}
    \begin{subfigure}[t]{0.32\textwidth}
        \centering
            \includegraphics[trim={1cm 0cm 2cm 0cm}, scale=0.32]{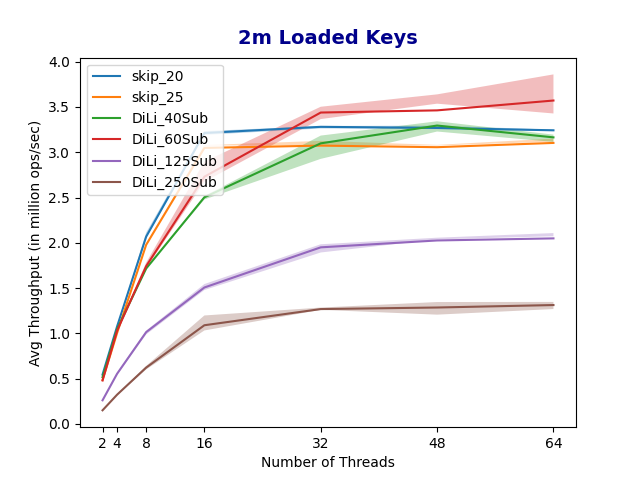}
            \caption{Maximum size throughput achieved when size of the list is 2m.}
            \label{fig:appSize2m}
    \end{subfigure}
    \begin{subfigure}[t]{0.32\textwidth}
        \centering
            \includegraphics[trim={1cm 0cm 2cm 0cm}, scale=0.32]{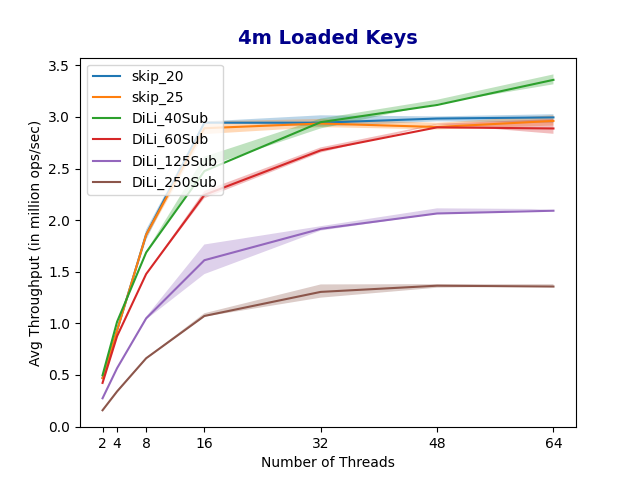}
            \caption{Maximum average throughput achieved when write proportion is 4m.}
            \label{fig:appSize4m}
    \end{subfigure}   
    \begin{subfigure}[t]{0.32\textwidth}
        \centering
            \includegraphics[trim={1cm 0cm 2cm 0cm}, scale=0.32]{Figures/library_size_workloads/Loaded_Total_5m_Max_Avg_Th.png}
            \caption{Maximum size throughput achieved when size of the list is 5m.}
            \label{fig:appWrite5m}
    \end{subfigure}   }
    \caption{Performance of DiLi and Skip List when used as a library to execute workloads of various key sizes. Write Proportion is maintained to be 50\% throughout.}
    \label{fig:AppSizeWorkloadPerformance}
\end{figure*}

    \Cref{fig:AppSizeWorkloadPerformance} shows the thread-wise results for all sizes that we have experimented.

\subsubsection{Practical Termination of Background Operations}
\label{subsec:appendixBackgroundOpTerminationResults}
\begin{figure}[!h]
    \centering
    \includegraphics[width=0.5\linewidth]{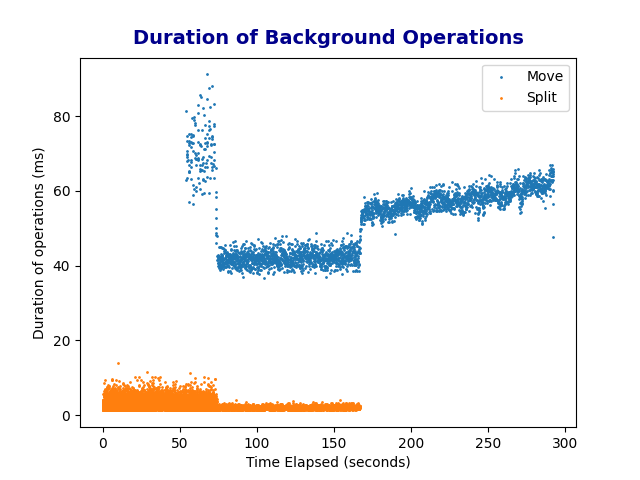}
    \caption{Scatter plot of the time taken by background operations, plotted against the local elapsed time at which the operation completed.}
    \label{fig:BackgroundOperations}
\end{figure}
\Cref{fig:BackgroundOperations} shows the time taken by the $\Split$ and $\Move$ operations ($\Move$ includes Switch) on the server that initially owned the entire key range for the experiment. The experiment was a workload of insert operations that loaded 1M keys into two 2-core machines, with one of them not owning any part of the list key range at the beginning, and begins to receive sublists to own through Move operations the first minute of the load test. We observe that the load of insertions completed at the 180 second mark, after which no Split operations were performed (since all sublists are now within the threshold limit of 125 and the list is no longer receiving updates). After this time, only Moves were performed to complete load balancing.
\section{Proof of Correctness for the Background Operation Algorithms}
\label{sec:AppendixProofs}

\subsubsection{Properties of \StartCount and \EndCount of a Sublist for Delegations}
\label{sec:AppendixCounters}

\begin{theorem}
    There is only one active subhead pointer to a sublist at any given time.
\end{theorem}
\begin{proof}
    For any sublist, a subhead pointer that stores an item with \StartCount as negative would redirect a request using either their \newLocation or a fresh registry lookup. However, during a Move, there will exist two subheads with \StartCount being non-negative -- \begin{enumerate*}
        \item The subhead pointer (SH) of the source machine sublist copy.
        \item The subhead pointer (SH*) created in the new machine sublist copy.
    \end{enumerate*}
    No registry entry stores SH* during the Move phase. Hence, every request gets routed through SH and will continue to treat it as the active subhead since its \StartCount is not negative. During a Switch, SH will have its \StartCount as negative, and thus reroute all requests from the outdated registry entry to the \newLocation, which is SH*. The $\Registry$ entry is then updated to SH*, after which rerouting via SH will no longer be required and SH* will be the only active subhead. 
\end{proof}

\begin{theorem}
    If an update on a sublist has checked that the sign of its \StartCount is positive after incrementing it, then the \StartCount of the sublist cannot become negative until the update also increments its \EndCount.
\end{theorem}
\begin{proof}
    When \StartCount increases without increasing \EndCount, the difference between the two is greater than \offset. This makes it impossible for the computed CAS value check (line \ref{line:minusiftyCAS}) to be true until \EndCount is also incremented. When \StartCount is already set to $-\infty$, no number of increments can make it positive, making the sublist copy inactive to perform.  
\end{proof}

    

\subsubsection{Bound on the Number of Delegations}
\label{sec:appendixDelegationCountBound}
\begin{theorem}
    The maximum number of network round trips involved in a client operation is 2 when there is no Switch operation, and 3 when there is a Switch operation. 
\end{theorem}
\begin{proof}
    When there is no Switch operation, the distributed sublist is static. There is exactly one machine that contains the sublist for the corresponding key. Either the client directly requests the corresponding machine (1 total round trip), or requests the machine that forwards this request (2 total round trips). Thus, the number of network round trips involved is at most 2 when there is no Switch operation.

    During a Switch operation, a request to the old machine gets delegated to the new machine, bringing about an additional network round trip to the computation. Such a request has to be served in the new machine, as the distributed sublist assumes that another Switch begins long after all existing requests to the old machine for the sublist are completed. Thus, the maximum number of network round trips of a client operation during a Switch is at most 3. 
\end{proof}

    Since every pair of round trip has a unique thread in common due to the nature of the delegation hops, the number of threads utilized per client operation is one more than the number of network round trips taken for the operation. In other words, the number of threads utilized is 3 when there is no Switch operation and 4 when there is a Switch operation.
\subsubsection{Correctness of the Replay Algorithm in Move}
\label{sec:AppendixReplay}

    \Cref{subsec:MoveSwitch} shares the intuition behind the replay. The replay algorithm to recreate the moving sublist is done by re-imagining the sublist to be built by $\InsertAfter$(prevItem, newItem), instead of Insert(key), as it offers the exact item at which the insertion took place. Note that prevItem and newItem have sId, ts, key members present inside them, for easy notation in the proof. This distinguishes multiple items with the same key that were entered and removed and ensures that if the prevItem had the same key as an old deleted item, then the replay waits until the newItem is inserted using the logical time stamp. Similarly, Remove(key) is replaced by $\Delete$(prevItem). 
    
    Upon receiving a request of the form RepInsert(prevItem, item, oldLocation), the operation is required to find the location of $prevItem$ using their unique (sID, ts) tuple and insert the $item$ at its appropriate place, and return the inserted location back to the requesting server. This server can now set the item that matches $oldLocation$ to have a $\newLocation$ provided from the replay message. While finding $prevItem$ and returning a item after insertion is straightforward, performing the insertion at the appropriate place requires some observations. Replaying a Delete is straightforward, as it only needs to find an item of a specific (sID, ts) and ensure that it is marked. 

        For the sake of this discussion, we say that if $\InsertAfter$(Ref(X), Y) has occurred, then Y is a \textit{successor} of X, and X is a \textit{predecessor} of Y. We say that X is an \textit{ancestor} of Y if and only if there is a sequence $X, X_1, X_2, \cdots, Y$ such that, each element in the list is a predecessor of the next element. In this case, we also say that $Y$ is a \textit{descendent} of $X$.
 For simplicity, the oldLocation parameter is hidden in the discussion, as it does not affect the replay and is only used to complete the assignment of a \newLocation on callback. We additionally use the terms list item and list node interchangeably.
    \begin{lemma}
    
    If two requests $\InsertAfter$ (A,B) and $\InsertAfter$ (A,C) occur simultaneously, then the node that gets inserted first will have a lower timestamp and will be farther away from A.\label{LemmaForTimestamp}
    \end{lemma}
    \begin{proof}
        
        The lower timestamp clause of the Lemma follows from the fact that the insertion that lost the race in CAS will re-increment the logical clock to get a timestamp higher than the previous values. The latter part of the Lemma follows from the fact that the item from the latter $\InsertAfter$ will occupy the next pointer of A and will have its own next pointer pointing to the element that was inserted at A earlier.
    \end{proof}
    
    \begin{lemma}
    If there is a replicate message of the form RepInsert(A,C), then $A.ts < C.ts$.
    \label{LemmaforReplicate}
    \end{lemma}
    \begin{proof}
    This follows from the fact that A must have existed in the \sublist before $\InsertAfter$ (Ref(A), C) was called.
    \end{proof}

    \begin{lemma}
        
    If there is a link of the form $A\rightarrow B$ in the \sublist that is being reconstructed, then one out of the following two has to be true: (1) node B was inserted at node A or descendants of A (and so has a timestamp greater than A); (2) B was inserted at a node preceding A (and so has a timestamp less than that of A).\label{LemmaForABeforeB}
    \end{lemma}
    \begin{proof}
    This follows from the fact that the entire linked list structure (excluding the next pointer mark values) has been created by the use of $\InsertAfter$ and $\Split$ operations. Since $\Split$ only inserts special, client-invisible nodes($\subhead$ and $\subtail$), it is safe to argue that nodes A and B must have been inserted into the list through $\InsertAfter$ operations. $\InsertAfter$ can only insert nodes to the right side of the node provided to it. Hence, if node B appears to the right of A, then either it was inserted at A or was inserted at an ancestor of A, at a time before A was inserted into the list. Thus, in the first case, $B.ts > A.ts$ and in the latter case, $B.ts < A.ts$.
    \end{proof}

    \begin{lemma}
    If the replicated server has a link of the form $A \rightarrow B$ and $A.ts > B.ts$, then any RepInsert(A,C) must insert C in a place after A, but before B.\label{LemmaB}
    \end{lemma}
    \begin{proof}
    This follows from \Cref{LemmaForABeforeB}, by which B was not inserted at A in this case. Hence B is not a competing insertion at A, and C can be inserted just after A as the first replicated insertion at A.
    \end{proof}

    \begin{lemma}
    If the replicated server has a link of the form $A\rightarrow B$, and $A.ts < B.ts$, then any RepInsert(A,C) must traverse from A and find a node D such that $C.ts > D.ts$ and insert C just before D.\label{LemmaD}
    \end{lemma}

    \begin{proof}        
    From \Cref{LemmaForABeforeB}, this conveys that B was inserted at A. This means C and B were possibly competing to insert at A, and so if $C.ts < B.ts$, C has to be present after B, according to \Cref{LemmaForTimestamp}. After B, there could exist other descendants of A that were inserted in ways similar to B. Hence, the replay has to find the first node D after A, such that $C.ts > D.ts$, which will ensure either that D is not a descendant of A or that D is a descendant of A that was inserted into the list before C was inserted. Either case, C has to be inserted just before D.
    \end{proof}

\begin{theorem}\label{theorem:TRReplay}
        The Replay algorithm reconstructs the \sublist with the exact structure that is present in the source machine of the $\Move$. 
    \end{theorem}
    \begin{proof}
    Combining  \Cref{LemmaB} and \Cref{LemmaD}, the algorithm to replay a request of RepInsert(prevItem,item) condenses down to finding the first node($curr$ in Line \ref{line:replayFindCurr} of \Cref{alg:MoveAndReplay}), such that $curr.ts < item.ts$ (as in Line \ref{line:replayTsCompare}). This will ensure that $curr$ is either a node of type B in \Cref{LemmaB} or a node of type D in \Cref{LemmaD}. Hence, the insert replays reconstruct the same \sublist that the corresponding $\InsertAfter$ operations had created on the other server. 
    \end{proof}

\subsubsection{Termination Condition of the Asynchronous Background Operations}
\label{subsec:appendixTerminationProof}
    While we will be showcasing the practical termination of the background operations in \Cref{sec:Empirical Evaluation}, and proved its asynchronous nature on client operations through lock-freedom, we still identify their termination conditions in this subsection.
    
    \textbf{Split:} While the correctness of Split is straightforward from the explanation in \Cref{subsec:split}, the termination of Split operation depends on overcoming three different race conditions -- \begin{enumerate*}
        \item It first competes with the insert operations (Line \ref{line:splitInsertion} on $sItem$ to insert the ``ST-SH'' block into the list.
        \item After insertion, it loops to find the new offsets for the split sublists (Line \ref{line:splitOffsetEnd}). This requires a brief moment in time when no updates take place on either sublist.
        \item Finally, it competes with other background Split operations on other sublists to update the registry (Lines \ref{line:splitRegistryStart} to \ref{line:splitRegistryEnd}).
    \end{enumerate*}

    \textbf{Move:} The correctness of Move comes from the exhaustive proof of correctness of the Replay Algorithm (provided in \Cref{sec:AppendixReplay}) to generate a live clone of the sublist. However, its termination depends on only one race condition - setting \StartCount to $-\infty$ through the CAS in Line \ref{line:minusiftyCAS} (succeeding this CAS initiates the Switch). This is a type of spin lock, that requires only a brief moment when there are no ongoing updates to the sublist. 

    \textbf{Switch:} The correctness of the switch comes from the single active subhead pointer property (as in \Cref{sec:AppendixCounters}). It, however, has one race condition -- It needs to find the latest subtail of the previous sublist (if one exists) and change its next pointer (Line \ref{line:switchSTStart} to \ref{line:switchSTEnd}). This thus competes with Move and Switch operations of that previous sublist, which is practically trivial to overcome as Move of the same sublist should anyway not be invoked too often to avoid performance loss from too many delegations of client operations on this sublist.

\end{document}